\documentclass[journal,10pt,twocolumn,twoside]{IEEEtran}
\usepackage{multicol}   
\usepackage{multirow}
\usepackage[pdftex]{graphicx}
\usepackage{algorithm,algpseudocode}
\usepackage[small]{caption}
\usepackage{float}
\usepackage{amssymb,amsmath}
\usepackage{graphicx}
\usepackage{subfigure}
\usepackage{xspace}
\usepackage{amsthm}
\usepackage{fancyhdr}

\usepackage{caption,setspace}
\usepackage{amssymb,amsmath}
\usepackage{bm}
\usepackage{textcomp} 
\usepackage{epstopdf}
\usepackage{bbding}
\usepackage{cite} 
\usepackage{color}
\usepackage{pifont}
\usepackage{lipsum}
\usepackage{amstext}
\usepackage{orcidlink}

\usepackage{cuted}

\graphicspath{{fig/}}

\newcommand\scalemath[2]{\scalebox{#1}{\mbox{\ensuremath{\displaystyle #2}}}}

\begin{document}

\title{\huge Robust Design of Beyond-Diagonal Reconfigurable Intelligent Surface Empowered RSMA-SWIPT System Under Channel Estimation Errors}
\author{Muhammad Asif \,\orcidlink{0000-0002-9699-1675}, Zain Ali \,\orcidlink{0000-0002-1880-8046}, Asim Ihsan \,\orcidlink{0000-0001-7491-7178}, Ali Ranjha \,\orcidlink{0000-0001-6663-3714}, Zhu Shoujin \,\orcidlink{0000-0002-2615-2489}, Manzoor Ahmed \,\orcidlink{0000-0002-0459-9845}, \IEEEmembership{Senior Member, IEEE}, Xingwang Li \,\orcidlink{0000-0002-0907-6517}, \IEEEmembership{Senior Member, IEEE}, and Symeon Chatzinotas \,\orcidlink{0000-0001-5122-0001}, \IEEEmembership{Fellow, IEEE}

\thanks{This work was supported in part by the Key Research and Development Projects in Tongling City under Grant No. 20230201014; and in part by the Tongling University Talent Research Initiation Fund Project under Grant No. 2023tlxyrc17.}

\thanks{Muhammad Asif and Zhu Shoujin are with the School of Electrical Engineering, Tongling University, Anhui, Tongling, China (e-mails: masif@tlu.edu.cn, 2023028@tlu.edu.cn).
	
Zain Ali is with the College of Science and Engineering, Hamad Bin Khalifa University, Qatar (e-mail: zainalihanan1@gmail.com)

Asim Ihsan is with the Department of Electrical Engineering, University of Cambridge, Cambridge, United Kingdom (e-mail: ai422@cam.ac.uk)

Ali Ranjha is with the École de Technologie Supérieure, Montréal, Quebec, Canada, (Email: ali-nawaz.ranjha.1@ens.etsmtl.ca).

Manzoor Ahmed is with the School of Computer and Information Science and the Institute for AI Industrial Technology Research, Hubei Engineering University, Xiaogan 432000, China (Email: manzoor.achakzai@gmail.com).
 
Xingwang Li is with the School of Physics and Electronic Information Engineering, Henan Polytechnic University, Jiaozuo, 454003, China (Email: lixingwangbupt@gmail.com).

Symeon Chatzinotas is with the Interdisciplinary Centre for Security, Reliability and Trust (SnT), University of Luxembourg, 1855 Luxembourg City, Luxembourg (e-mail: symeon.chatzinotas@uni.lu).

(Corresponding author: Zhu Shoujin)
}

\vspace{-0.6cm}}%

\markboth{}
{ \MakeLowercase{\textit{}}} 
\maketitle

\begin{abstract}
This work explores the integration of rate-splitting multiple access (RSMA), simultaneous wireless information and power transfer (SWIPT), and beyond-diagonal reconfigurable intelligent surface (BD-RIS) to enhance the spectral-efficiency, energy-efficiency, coverage, and connectivity of future sixth-generation (6G) communication networks. Specifically, with a multiuser BD-RIS-empowered RSMA-SWIPT system, we jointly optimize the transmit precoding vectors, the common rate proportion of users, the power-splitting ratios, and scattering matrix of BD-RIS node, under the assumption of imperfect channel state information (CSI). Additionally, to better capture practical hardware behavior, we incorporate a nonlinear energy harvesting model and ensure that the resulting system satisfies all energy harvesting constraints. In the considered system, we design a robust optimization framework to maximize the system sum-rate, while explicitly accounting for the worst-case impact of CSI uncertainties. To tackle the inherent non-convexity of the problem, we introduce an alternating optimization framework that partitions the problem into several blocks, which are optimized in an iterative manner. More specifically, the transmit precoding vectors are optimized by reformulating the problem as a convex semidefinite programming through successive-convex approximation (SCA), whereas the inherently convex power-splitting problem is solved using the MOSEK-enabled CVX toolbox. Subsequently, to optimize the scattering matrix of the BD-RIS, we first employ SCA to reformulate the problem into a convex form, and then design a manifold optimization strategy based on the Conjugate-Gradient method. Finally, numerical simulations are conducted to evaluate the performance of the proposed scheme, revealing significant performance improvements over existing benchmarks and demonstrating rapid convergence within a reasonable number of iterations.
\end{abstract}


\IEEEpeerreviewmaketitle

\section{Introduction}
\IEEEPARstart {T}{he} deployment of fifth-generation (5G) mobile communications networks has driven significant progress in diverse areas, enabling innovative applications such as autonomous vehicles, remote healthcare, and smart urban infrastructure \cite{wang2022gcwcn, zhang20196g}. Despite these advancements, the rapid emergence of data-intensive and mission-critical services has exposed fundamental limitations of 5G systems, particularly concerning network capacity, coverage, energy-efficiency, and robustness under challenging conditions  \cite{saad2019vision}. Motivated by these challenges, researchers and industry stakeholders are actively engaged in conceptualizing and developing sixth-generation (6G) communication networks, aiming to address the growing need for ubiquitous and intelligent connectivity capable of supporting the unprecedented requirements of next-generation applications and services \cite{chen2020vision, wang2023road}. To achieve the aforementioned ambitious goals envisioned for 6G, breakthrough technologies such as rate-splitting multiple access (RSMA) \cite{mao2022rate, clerckx2023primer}, simultaneous wireless information and power transfer (SWIPT) \cite{perera2017simultaneous}, and reconfigurable intelligent surface (RIS) \cite{di2020smart} are being explored to significantly enhance spectral-efficiency, energy-efficiency, coverage, and connectivity in future wireless networks.

In this regard, RSMA is an efficient physical-layer technology for non-orthogonal transmission that provides enhanced interference management capability compared to conventional multiple access techniques \cite{mao2022rate}. In the RSMA protocol, the transmitter divides each user’s message into a common part and a private part. The private part is individually encoded into a private stream using a user-specific private codebook, whereas the common messages intended for multiple users are combined and encoded into one unified stream using a publicly available codebook shared among all users. On the receiver end, each user initially decodes the common stream and then employs successive interference cancellation (SIC) to eliminate the common message from the received signal. After this step, each user proceeds to decode its own private stream while treating the private signals of other users as interference  \cite{clerckx2023primer}. Thus, RSMA stands out as a versatile multiple access strategy that strikes a balance between two extremes: Unlike space-division multiple access (SDMA), which considers interference as noise, and non-orthogonal multiple access (NOMA), which decodes all interfering signals entirely, it adopts a partial decoding strategy for interference.\cite{katwe2022rate}. As a result, the balanced strategy adopted by RSMA enhances spectral-efficiency and robustness in multi-user systems. Additionally, given the energy consumption demands of billions of connected devices in future 6G networks, wireless energy transmission emerges as a sustainable solution for enabling ubiquitous connectivity \cite{lopez2021massive}. In this context, SWIPT provides a flexible and efficient means to deliver continuous and reliable energy alongside data transmission \cite{xu2019robust}. In particular, in SWIPT-enabled systems, the received RF signal at each user terminal is partitioned into two streams: one dedicated to information decoding and the other to energy harvesting. This division is typically implemented using either power-splitting (PS) or time-switching (TS) modes, with the PS mode generally achieving a superior trade-off between data rate and energy-efficiency \cite{xu2019robust}. Thus, the integration of RSMA with the SWIPT protocol presents a promising solution to simultaneously address the energy and spectral-efficiency requirements essential for realizing future 6G communication networks. 

Meanwhile, RIS has attracted considerable attention for its flexible capability to dynamically manipulate the wireless propagation environment, thereby enhancing the energy-efficiency, coverage, and connectivity of wireless networks \cite{di2020smart,asif2024securing}. Typically, an RIS is a planar surface composed of a large number of passive reflective elements that can redirect incoming signals toward desired directions by dynamically modifying their amplitude and phase shifts \cite{ihsan2022energy}. Inspired by these advantages of RIS, several researchers are investigating its potential benefits by integrating RIS with RSMA and SWIPT technologies 
\cite{zhang2023energy,asif2024leveraging,camana2022rate,hashempour2024secure,van2024swipt}. Specifically, the authors in \cite{zhang2023energy} proposed a deep reinforcement learning-based optimization framework to maximize the energy-efficiency of an RIS-empowered RSMA-SWIPT system, while considering power budget and quality-of-service (QoS) requirements. In \cite{asif2024leveraging}, the authors proposed an intelligent resource allocation strategy to enhance the performance of a simultaneously transmitting and reflecting reconfigurable intelligent surface (STAR-RIS)-enabled RSMA-SWIPT system under hardware impairments at both the transmitter and receiver. To minimize the transmission power of the base station (BS), an efficient resource management algorithm was proposed for an RIS-assisted RSMA-SWIPT system under imperfect channel state information (CSI) \cite{camana2022rate}. The authors in \cite{hashempour2024secure} proposed an optimization algorithm to maximize the achievable worst-case sum secrecy-rate of an RIS-assisted RSMA-SWIPT system under imperfect CSI for untrusted receivers. Furthermore, the authors in \cite{van2024swipt} investigated an optimization framework to enhance the performance of an RIS-empowered RSMA-based ad hoc network integrated with a SWIPT architecture, considering both PS and TS protocols.

Nevertheless, the aforementioned studies\cite{zhang2023energy,asif2024leveraging,camana2022rate,hashempour2024secure,van2024swipt} focus exclusively on the conventional RIS architecture, also known as single-connected RIS, in which the scattering matrix is represented as a diagonal phase-shift matrix. Specifically, each RIS element operates independently without interconnections with other elements, thus only controlling the phase of incoming signals through the diagonal entries of the scattering matrix. However, a new RIS architecture known as beyond diagonal RIS (BD-RIS) has recently emerged, which employs a full scattering matrix that is not limited to diagonal elements and can dynamically adjust the impedance between elements. This design enables BD-RIS to simultaneously manipulate both the phase and amplitude of incident waves, thereby enhancing system performance relative to conventional RIS architectures \cite{shen2021modeling,li2022beyond}. In this context, researchers have proposed several resource allocation strategies based on the BD-RIS architecture to improve the performance of different wireless communication networks. For instance, the authors in \cite{qin2025joint} proposed a joint resource optimization framework for a BD-RIS-enabled cooperative mobile-edge computing system, aiming to maximize the total number of completed task bits by jointly optimizing the CPU frequencies, BD-RIS scattering matrix, bandwidth allocation, transmit power, and energy transfer time. In \cite{zhou2023optimizing}, an intelligent optimization strategy was proposed to maximize energy-efficiency and minimize transmission power by adjusting the scattering matrix of the BD-RIS while satisfying the system's QoS requirements. Further, the authors in \cite{li2022reconfigurable} proposed a non-reciprocal BD-RIS architecture characterized by a non-diagonal scattering matrix, where sum-rate maximization was achieved by jointly optimizing the active beamforming and the BD-RIS scattering matrix. Moreover, an efficient channel estimation and beamforming technique was proposed for BD-RIS-assisted multi-antenna systems \cite{li2024channel}. In \cite{zhang2025sum}, the authors proposed an efficient resource-allocation framework to maximize the achievable sum-rate in a BD-RIS-assisted transportation system under a multi-cell scenario, by jointly optimizing power allocation and the BD-RIS scattering matrix.

However, all of the aforementioned studies \cite{qin2025joint,zhou2023optimizing,li2022reconfigurable,li2024channel,zhang2025sum} proposed resource allocation strategies that focused exclusively on the BD-RIS architecture, without synergizing RSMA and SWIPT technologies. In this regard, the authors in \cite{soleymani2023optimization} proposed a general optimization framework for RSMA-enabled BD-RIS-assisted ultra-reliable low-latency communications (URLLC) system, in which both the objective function and the constraints are formulated as linear functions of the achievable rates and energy-efficiency. In \cite{de2025robustness}, the authors proposed two interference attack strategies, random and aligned, against RSMA-based multi-user multiple-input multiple-output (MIMO) systems with BD-RIS architectures. They designed random reflections using Takagi factorization and formulated aligned attacks through quadratically constrained quadratic program (QCQP) based optimization to systematically disrupt channel acquisition induced by BD-RIS. Accordingly, the authors in \cite{huroon2023optimized} developed an alternating optimization framework aimed at maximizing the sum-rate of an RSMA-based BD-RIS-enabled unmanned aerial vehicle (UAV) network. Finally, the authors in \cite{li2024synergizing} proposed an optimization approach to maximize the average sum-rate of an RSMA-assisted BD-RIS system under imperfect CSI, by jointly optimizing the transmit precoding vectors and the scattering matrix of the BD-RIS. However, the existing studies \cite{soleymani2023optimization, de2025robustness, huroon2023optimized, li2024synergizing} have the following limitations: 1) Firstly, except for \cite{li2024synergizing}, none considered the impact of imperfect CSI. Since it is often unrealistic to obtain perfect CSI for all links in challenging environments, the optimization frameworks proposed in these works lack robustness; 2) Secondly, as mentioned above, synergizing RSMA and BD-RIS with SWIPT technology presents a promising solution to simultaneously address the energy and spectral-efficiency requirements essential for future 6G applications. However, none of these studies considered integrating SWIPT into their proposed frameworks; 3) Thirdly, all existing studies \cite{soleymani2023optimization, de2025robustness, huroon2023optimized, li2024synergizing} focused solely on enforcing the unitary constraint when computing the BD-RIS scattering matrix, while ignoring critical system constraints such as users’ QoS constraints during the passive beamforming optimization process, primarily to reduce the complexity of problem. Although these constraints are often incorporated when optimizing other variables, neglecting them in the passive beamforming stage may result in solutions that violate essential system requirements.

Motivated by the aforementioned limitations in existing works, we propose a robust resource-allocation framework for a BD-RIS-assisted RSMA-SWIPT system, which accounts for imperfect CSI in both the direct and cascaded channels, while simultaneously satisfying QoS, power budget, common message decoding, energy harvesting, and unitary constraints. Accordingly, the key contributions of this work are highlighted below:
\begin{itemize}
\item We consider an RSMA-based BD-RIS-assisted SWIPT system in which users employ a PS architecture to divide the received signal power into two parts: one for information-decoding and the other for energy-harvesting. To better capture practical behavior, a non-linear energy harvesting model is employed while ensuring that the system’s energy-harvesting constraints are satisfied. 

\item To enhance the robustness of the considered system, the sum-rate maximization is carried out under imperfect CSI at the BS by adopting a bounded channel estimation error model that accounts for the worst-case effects of CSI uncertainties.

\item Next, we propose an intelligent resource optimization strategy to maximize the system’s achievable sum-rate by jointly optimizing several key parameters, including the transmit precoding vectors of the common and private streams, the power-splitting ratios, the common rate proportions of users, and the scattering matrix of the BD-RIS, while satisfying QoS, power budget, energy harvesting, and unitary constraints.

\item Moreover, by leveraging the successive convex approximation (SCA) technique, the optimization problem for designing the transmit beamforming vectors of both the common and private streams is reformulated as a convex semi-definite programming (SDP) problem, which is then solved using the Mosek-enabled CVX toolkit. Further, the inherently convex optimization problem for computing power-splitting ratios is solved using Mosek-assisted CVX toolbox. Likewise, to compute the scattering matrix of the BD-RIS, we first apply the SCA technique to transform the original problem into a convex form. Subsequently, we develop a manifold optimization strategy based on the Conjugate-Gradient method, where the computation of the Riemannian gradient involves projecting the standard Euclidean gradient onto the manifold’s tangent space.
\item  Unlike the existing works that overlook critical system constraints, particularly those related to the minimum rate requirements for satisfying the QoS of multiple users, our approach explicitly incorporates these constraints into the BD-RIS beamforming optimization. This ensures that the resulting solution remains feasible and meets all system requirements.

\item Finally, the proposed scheme is evaluated through numerical simulations, which demonstrate its notable performance gains compared to benchmark methods, along with fast convergence achieved in a limited number of iterations.

\end{itemize}
   \begin{figure}[t]
  	\centering
  	\includegraphics [width=0.35\textwidth]{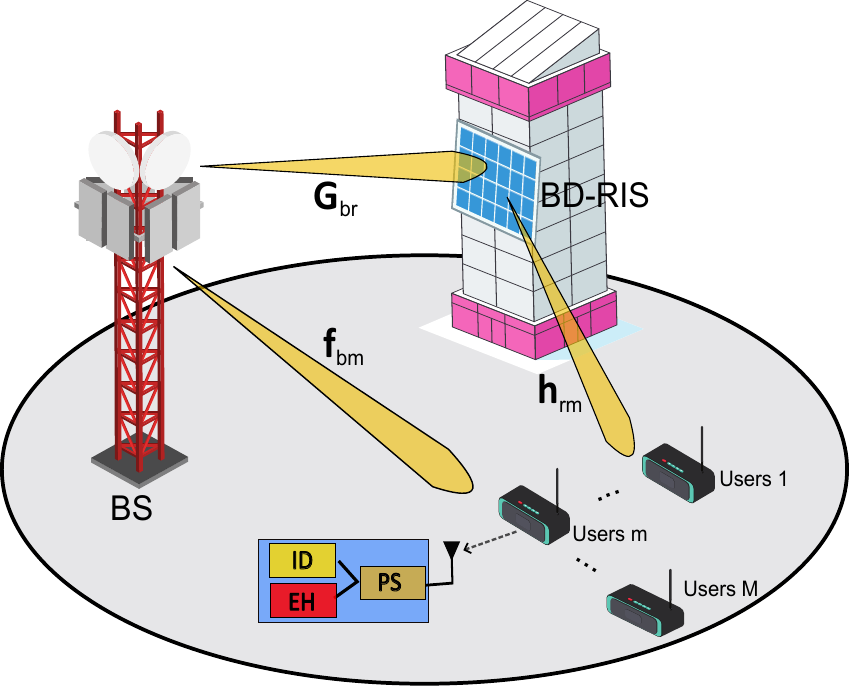}
  	\caption{Illustration of system model}
  	\label{f1}
  \end{figure}           
\section{System Model and Problem Formulation}
We consider a system setup in which a BS equipped with $K$ antennas, configured in a uniform linear array, serves $M$ single-antenna devices, each employing a power-splitting (PS) architecture that splits the received signal into two parts dedicated to energy harvesting (EH) and information decoding (ID) \cite{xu2019robust}. Operating under the RSMA protocol, the system integrates a BD-RIS comprising $L$ scattering elements configured as a uniform rectangular array to serve RSMA users. The channel from BS to BD-RIS is denoted as $\mathbf{G}_{br} \in \mathbb{C}^{L \times K}$, the channel between BD-RIS and $m^{th}$ user is denoted as  $\mathbf{h}_{rm} \in \mathbb{C}^{L \times 1}$, and the direct channel between the BS and $m^{\text{th}}$ RSMA user is denoted as $\mathbf{f}_{bm} \in \mathbb{C}^{K \times 1}$. Additionally, it is assumed that all of these channels,  $\mathbf{G}_{br}$, $\mathbf{h}_{rm}$, $\mathbf{f}_{bm}$, $\forall m\in \mathcal M = \{1,2, ..., M\}$, are subject to Rician fading model. The Rician fading model for any channel $\tilde{\kappa} \in \{\mathbf{G}_{br}, \mathbf{h}_{rm}, \mathbf{f}_{bm}\}$ can be expressed as
\begin{equation}
	\tilde{\kappa} = \sqrt{\tilde{\varrho}_{i}} \left( \sqrt{\frac{\zeta_{i}}{1+\zeta_{i}}} \tilde{\kappa}_{\mathrm{LOS}} + \sqrt{\frac{1}{1+\zeta_{i}}} \tilde{\kappa}_{\mathrm{NLOS}} \right),
	\label{eq:rician_general}
\end{equation}
where $\zeta_{i}$ and $\tilde{\varrho}_{i}, \forall i \in \{br, rm, bm\}$, denote the Rician factor and the large-scale pathloss associated with the channel $\tilde{\kappa}$, respectively. Additionally, $\tilde{\kappa}_{\mathrm{LOS}}$ and $\tilde{\kappa}_{\mathrm{NLOS}}$ denote the line-of-sight (LOS) and non-LOS (NLOS) components of $	\tilde{\kappa}$, respectively.

Further, by employing the RSMA protocol, the signal transmitted by the BS node can then be expressed as  
\begin{align}
	&\mathbf x= \mathbf{v}_{0}w_{0} + {\sum\limits_{{\substack{m=1}}}^{M} \mathbf{v}_{m}w_{m} , \forall m\in \mathcal M},\label{2}
\end{align}
where $\mathbf{v}_{0}$ and $\mathbf{v}_{m}$ denote the transmit precoding vectors for the common and private data streams of RSMA users, respectively. $w_{0}$ and $w_{m}$ denote the data signals corresponding to the common and private messages, respectively. Additionally, let $\mathbf {\Theta} \in \mathbb{C}^{L \times L}$ denotes the full-rank scattering matrix for BD-RIS. Then, the signal received at the $m^{\text{th}}$ user served under RSMA protocol can be expressed as 
\begin{align}
	& y_{m}= \mathbf {h}_m \mathbf x + \tilde{n}_{m}, \forall m\in \mathcal M,\label{3}
\end{align}
where $\mathbf {h}_m= \mathbf{f}_{bm} + \mathbf{G}^H_{br}\mathbf {\Theta} \mathbf{h}_{rm}$ denotes the equivalent channel gain from BS to $m^{\text{th}}$ user. Further, $\tilde{n}_{m}$ denotes zero-mean additive noise with variance $\tilde{\sigma}_{m}^{2}$. Further, based on the SWIPT protocol, the signals received at the ID and EH circuits are expressed as follows
\begin{align}
	& y^{dec.}_{m}= \sqrt{\beta_m}\Big( \mathbf {h}_m \mathbf x + \tilde{n}_{m} \Big) + \tilde{n}_{dec.}, \label{4}
\end{align} 
and,
\begin{align}
	& y^{har.}_{m}= \sqrt{1-\beta_m}\Big( \mathbf {h}_m \mathbf x + \tilde{n}_{m} \Big). \label{5}
\end{align}
where $\tilde{n}_{\mathrm{dec.}} \sim \mathcal{CN}\big(0, \tilde{\sigma}^2_{\mathrm{dec.}}\big)$ represents the additive complex Gaussian noise introduced by the ID circuit.

Further, it is important to highlight that the power-splitting architecture at each user splits the power of incoming radio signal based on the power-splitting ratio $\beta_m \in (0,1)$ \cite{camana2022rate}. Thus, the instantaneous rate at the ID circuit of $m^{\text{th}}$  user to decode the common signal can be written as
\begin{align}
	R^o_{m}= & \log_{2}\left(1+\frac{ \beta_m |\mathbf{v}^H_{0}\mathbf h_{m}|^2 }  {\beta_m \Bigl( \sum\limits_{{\substack{i=1}}}^{M} |\mathbf h_{m} \mathbf{v}_{i}|^2 + \tilde{\sigma}_{m}^{2}  \Bigr)+\tilde{\sigma}_{dec.}^{2} } \right) . \label{6}
\end{align}

Once the common stream is correctly decoded, the receiver employs SIC to eliminate it from the received signal. Subsequently, each RSMA user decodes its private message by treating the private messages intended for other users as interference. Thus, the instantaneous rate desired at the ID circuit to decode the private message can be expressed as
\begin{align}
	R^p_{m}= & \log_{2}\left(1+\frac{ \beta_m |\mathbf{v}^H_{m}\mathbf h_{m}|^2 }  {\beta_m \Bigl( \sum\limits_{{\substack{i=1 \\ i\neq m}}}^{M} |\mathbf h_{m} \mathbf{v}_{i}|^2 + \tilde{\sigma}_{m}^{2}  \Bigr)+\tilde{\sigma}_{dec.}^{2} } \right) . \label{7}
\end{align}

Further, we consider that the CSI of equivalent channel $\mathbf h_{m}$ can not be obtained perfectly. Thus, under the assumption of imperfect CSI, we model the equivalent channel as $\overline{\mathbf{h}}_{m}=\mathbf h_{m} + \Delta \mathbf h_{m}$, where $\mathbf h_{m}$ and  $\Delta \mathbf h_{m}$ denote the acquired CSI and channel estimation error with complex Gaussian distribution with zero mean and variance $\sigma_{\Delta_k}^{2}$  \cite{li2022robust}, respectively. Further, we assume that a certain threshold for $\Delta \mathbf h_{m}$ can be obtained as $\left\|\Delta \mathbf{h}_m\right\| \leq \delta_m$ \cite{li2022robust}. Assuming that $\overline{\mathbf{h}}_{m}$ and $ \Delta \mathbf h_{m}$ are independent of each other, the channel covariance matrices are expressed as $\overline{\mathbf{H}}_{m}=\mathbf H_{m} + \Delta \mathbf H_{m}$, where $\overline {\mathbf{H}}_m=\mathbb{E}\left\{\overline{\mathbf{h}}_{m} \overline{\mathbf{h}}_{m}^H\right\}$ and $\mathbf{H}_m=\mathbb{E}\left\{\mathbf{h}_m \mathbf{h}_m^H\right\}$ denote the covariance matrices of real channel and estimated equivalent channel, respectively. Additionally, $ \Delta \mathbf{H}_m=\mathbb{E}\left\{\Delta \mathbf{h}_m \Delta \mathbf{h}_m^H\right\}$ denotes the covariance uncertainty matrix.

Thus, under imperfect CSI, \eqref{6} and \eqref{7} can be updated as follows:
\begin{align}
	\overline{R}^o_{m}= & \log_{2}\left(1+\frac{ \beta_m \mathrm{tr} \Big(\mathbf{v}_{0}^H \big(\mathbf H_{m} + \Delta \mathbf H_{m}\big) \mathbf{v}_{0} \Big) }  {\beta_m  \sum\limits_{{\substack{i=1}}}^{M} \mathrm{tr} \Big(\mathbf{v}_{i}^H \big(\mathbf H_{m} + \Delta \mathbf H_{m}\big) \mathbf{v}_{i} \Big) + \overline{\sigma}_{m}^{2}} \right) , \label{8}
\end{align}
\begin{align}
	\overline{R}^p_{m}= & \log_{2}\left(1+\frac{ \beta_m \mathrm{tr} \Big(\mathbf{v}_{m}^H \big(\mathbf H_{m} + \Delta \mathbf H_{m}\big) \mathbf{v}_{m} \Big) }  {\beta_m  \sum\limits_{{\substack{i=1 \\ i\neq m}}}^{M} \mathrm{tr} \Big(\mathbf{v}_{i}^H \big(\mathbf H_{m} + \Delta \mathbf H_{m}\big) \mathbf{v}_{i} \Big) + \overline{\sigma}_{m}^{2}} \right), \label{9}
\end{align}
where $\overline{\sigma}_{m}^{2} = \beta_m \tilde{\sigma}_{m}^{2}  +\tilde{\sigma}_{dec.}^{2}$. Further, for successful decoding of common data stream, the following condition must be satisfied as 
\begin{align}
\sum\limits_{{\substack{m \in \mathcal M}}} \bar{r}_m^c	\leq \log_{2}\left(1+\frac{ \beta_m \mathrm{tr} \Big(\mathbf{v}_{0}^H \big(\mathbf H_{m} + \Delta \mathbf H_{m}\big) \mathbf{v}_{0} \Big) }  {\beta_m  \sum\limits_{{\substack{i=1}}}^{M} \mathrm{tr} \Big(\mathbf{v}_{i}^H \big(\mathbf H_{m} + \Delta \mathbf H_{m}\big) \mathbf{v}_{i} \Big) + \overline{\sigma}_{m}^{2}} \right), \label{10}
\end{align}
where $\bar{r}_m^c$ corresponds to the common rate fraction allocated to the $m^{\text{th}}$ user. Accordingly, the sum-rate of the system can be written as
	\begin{align}
	\overline{R}_{sum}= & \sum\limits_{{\substack{m \in \mathcal M}}} \Big(\overline{r}^c_m + \overline{R}^p_{m}\Big), \forall m\in \mathcal M.  \label{11}
\end{align}

Furthermore, rather than relying on the traditional linear energy harvesting model, we adopt a more realistic non-linear energy harvesting model \cite{camana2022rate, xiong2017rate}, which is defined as follows:
\begin{align}
	\overline{E}_{m} = \frac{\Lambda(\tilde{E}_{m}) - \overline{C}_m \overline{D}_m}{1 - \overline{D}_m}, \quad \forall m \in \mathcal{M}, \label{12}
\end{align}
with
\begin{align}
		\tilde{E}_{m} = (1 - \beta_m) \Big( \sum_{i=0}^{M} \mathrm{tr}\bigl(\mathbf{v}_{i}^H (\mathbf{H}_{m} + \Delta \mathbf{H}_{m}) \mathbf{v}_{i}\bigr) + \tilde{\sigma}_{m}^{2} \Big). \label{13}
\end{align}
Here, $\overline{D}_m=\frac{1}{1+e^{\overline{c}_m \overline{d}_m}}$, $\Lambda(\tilde{E}_{m})=\frac{\overline{C}_m}{1+e^{-\overline{c}_m(\tilde{E}_{m}-\overline{d}_m)}}$, and $\overline{C}_m$ denotes the maximum power harvested by the energy harvesting circuit. Additionally, $\overline{c}_m$ and $\overline{d}_m$ are constants determined by the circuit characteristics \cite{camana2022rate, xiong2017rate}.

Next, the optimization problem for the considered system can be written as follows
\begin{subequations}\label{P1}
	\begin{align}
		\text{\textbf{(P1)}}	& \mathop {\max }\limits_{( \mathbf {v}_{0}, \mathbf {v}_{m},\overline{r}^c_m, \mathbf \Theta,  \beta_m)} \    \sum\limits_{{\substack{m=1}}}^M \Big(\overline{r}^c_m + \overline{R}^p_m\Big), \label{1114a}  \\
		s.t.\ C_1:\ & \overline{r}^c_m + \overline{R}^p_m  \geq  {R}_{min}, \forall m\in \mathcal M, \label{1114b} \\
		\ C_2:\ &  \sum\limits_{{\substack{m=1}}}^M  \overline{r}^c_m  \leq \overline{R}^o_m, \forall m\in \mathcal M, \label{1114c}\\
        \ C_3:\ & \left\|\mathbf v_0\right\|^2+\sum\limits_{{\substack{m=1}}}^M \left\|\mathbf v_m\right\|^2 \leq P_{max.}, \label{1114d}\\
        \ C_4:\ & \Lambda(\tilde{E}_{m})- \overline{C}_m \overline{D}_m \geq \vartheta_m(1-\overline{D}_m), \label{1114e} \\ 
        \ C_5: \ &0 \leq\beta_m \leq 1, \forall m\in \mathcal M, \label{1114f} \\
		\ C_6: \ &\mathbf \Theta \mathbf \Theta^H= \mathbf{I}_{L}, \label{1114g} 
	\end{align}
\end{subequations}
where ${R}_{min}$ denotes the minimum rate requirement to meet the QoS demand for each users, $\vartheta_m$ is the desired threshold for harvested energy. Additionally, Constraints $C_1$ and $C_2$ ensure the QoS requirement and successful decoding of common data stream for each user, respectively. Constraints $C_3$ and $C_4$ fulfill the power-budget and energy-harvesting requirements of the system. Moreover, it is important to emphasize that although the BD-RIS elements are interconnected, they cannot change the overall amplitude of the incoming radio signals. Instead, they merely redistribute the incident energy across the elements by routing power from one element to another, ensuring that the total reflected power remains equal to the incident power. Consequently, the BD-RIS's scattering matrix $\mathbf \Theta$ must satisfy the unitary constraint specified by $C_6$.  
%%%%%%%%%%%%%%%%%%%%%%%%%%%%%%%%%%%%%%%%%%%%%%%%%%%%%%%%%%%%%%%%%%%%%%%%%%%%%%%%%%%%%%%%%%%%%%%%%%%%%%%%
\section{Proposed Solution for the Optimization Problem}
Since the optimization problem \textbf{(P1)} is highly complex in nature, we first transform it into a tractable form to solve it. Next, we define \textbf{Lemma 1} as follows:\\
\textbf{Lemma 1:} For any Hermitian matrices $\mathbf{Y}$ and $\mathbf{Z}$, with $\mathbf{Y}$ being a rank-one positive semi-definite (PSD) matrix, if $\mathbf{Z}$ satisfies $|\mathbf{Z}| \leq \delta^{2}$, then
\begin{equation}
	\max _{\|\mathbf{Z}\| \leq \delta^2} \operatorname{tr}(\mathbf{Y Z})=\delta^2 \operatorname{tr}(\mathbf{Y}). \label{15}
\end{equation} 
\noindent \textbf{Proof:} The complete proof can be found in Appendix A.

Then, using the equalities $\operatorname{tr}(\mathbf v_m^H \Delta \mathbf H_m \mathbf v_m) = \operatorname{tr}(\Delta \mathbf H_m \mathbf V_m)$ 
and $\operatorname{tr}(\mathbf v_0^H \Delta \mathbf H_m \mathbf v_0) = \operatorname{tr}(\Delta \mathbf H_m \mathbf V_0)$, where $\mathbf V_m = \mathbf v_m \mathbf v_m^H$ and $\mathbf V_0 = \mathbf v_0 \mathbf v_0^H$ 
are PSD matrices satisfying $\operatorname{rank} (\mathbf V_m)=1$ and $\operatorname{rank} (\mathbf V_0)=1$. By exploiting \textbf{Lemma 1}, we can write as:
\begin{equation}
	\max _{\|\Delta \mathbf H_m\| \leq \delta_m^2} \operatorname{tr}(\Delta \mathbf H_m \mathbf V_m)=\delta_m^2 \operatorname{tr}(\mathbf{V_m}). \label{16}
\end{equation} 
and,
\begin{equation}
	\max _{\|\Delta \mathbf H_m\| \leq \delta^2} \operatorname{tr}(\Delta \mathbf H_m \mathbf V_0)=\delta_m^2 \operatorname{tr}(\mathbf{V_0}). \label{17}
\end{equation} 

Additionally, we consider a worst-case scenario in which the desired signal channels are assumed to be overestimated; thus, $\Delta \mathbf H_m$ needs to attain its minimum value of $-\delta_m^2 \mathbf I$, while the interference channels are underestimated; thus, $\Delta \mathbf H_m$ needs to reach its maximum value of $\delta_m^2 \mathbf I$ \cite{li2022robust,zheng2023zero}.

Thus, under the worst-case scenario, Eqs.~\eqref{8} and \eqref{9} can be updated as follows:
\begin{align}
	\tilde{R}^o_{m}= & 	 \log_{2}\left(1+\frac{ \beta_m \mathrm{tr} \Big(\big(\mathbf H_{m} - \delta_m^2 \mathbf I\big)\mathbf V_0\Big) }  {\beta_m  \sum\limits_{{\substack{i=1}}}^{M} \mathrm{tr} \Big(\big(\mathbf H_{m} + \delta_m^2 \mathbf I\big)\mathbf V_i\Big) + \overline{\sigma}_{m}^{2}} \right) , \label{18}
\end{align}
\begin{align}
	\tilde{R}^p_{m}= & 	\log_{2}\left(1+\frac{ \beta_m \mathrm{tr} \Big(\big(\mathbf H_{m} - \delta_m^2 \mathbf I\big)\mathbf V_m\Big) }  {\beta_m  \sum\limits_{{\substack{i=1 \\ i\neq m}}}^{M} \mathrm{tr} \Big(\big(\mathbf H_{m} + \delta_m^2 \mathbf I\big) \mathbf{V}_{i} \Big) + \overline{\sigma}_{m}^{2}} \right), \label{19}
\end{align}

Accordingly, Eq.~\eqref{13} can be updated as
\begin{align}
	\hat{E}_{m} = (1 - \beta_m) \Bigl( \sum_{i=0}^{M} \mathrm{tr}\bigl((\mathbf{H}_{m} - \delta_m^2 \mathbf I) \mathbf{V}_{i}\bigr) + \tilde{\sigma}_{m}^{2} \Bigr). \label{20}
\end{align}

As a result, we can update Eq.~\eqref{1114e} as follows 
\begin{align}
\vartheta_m(1-\overline{D}_m) \leq \Lambda(\hat{E}_{m})- \overline{C}_m \overline{D}_m.  \label{21}
\end{align}

As a result, the reformulated optimization problem can be written as
\begin{subequations}\label{P2}
	\begin{align}
		\text{\textbf{(P2)}}	& \mathop {\max }\limits_{( \mathbf {V}_{0}, \mathbf {V}_{m},\overline{r}^c_m, \mathbf \Theta,  \beta_m)} \    \sum\limits_{{\substack{m=1}}}^M \Big(\overline{r}^c_m + \tilde{R}^p_m\Big), \label{22a}  \\
		s.t.\ \ & \overline{r}^c_m + \tilde{R}^p_m  \geq {R}_{min}, \forall m\in \mathcal M, \label{22b} \\
		\ \ & \sum\limits_{{\substack{m=1}}}^M  \overline{r}^c_m  \leq \tilde{R}^o_m, \forall m\in \mathcal M, \label{22c}\\
		\ & \operatorname{tr}(\mathbf V_0)+\sum\limits_{{\substack{m=1}}}^M \operatorname{tr}(\mathbf V_m) \leq P_{max.}, \label{22d}\\
    	\ \ & \mathbf V_m \succeq 0, \ \mathbf V_0 \succeq 0, \ \forall m\in \mathcal M, \label{22e}\\ 
    	\ \ & \operatorname{rank}(\mathbf V_m)=1, \ \operatorname{rank}(\mathbf V_0)=1,  \forall m\in \mathcal M. \label{22f}\\
		\  \ & \eqref{1114f}, \eqref{1114g}, \eqref{21}. \label{22g} 
	\end{align}
\end{subequations}

Next, it is important to highlight that problem \textbf{(P2)} presents significant complexity and non-convexity, owing to both its non-convex objective function and non-convex constraints, which arise from the interdependence among the optimization variables. To tackle this, we adopt an alternating optimization strategy that partitions the original problem into multiple blocks, each of which is optimized iteratively.

\subsection{Optimization of $\mathbf v_m$ and $\mathbf v_0$: Stage 1}
To optimize the transmit beamforming vectors $\mathbf v_m$ and $\mathbf v_0$, the associated problem is formulated as follows:
\begin{subequations}\label{P3}
	\begin{align}
		\text{\textbf{(P3)}}	& \mathop {\max }\limits_{( \mathbf {V}_{0}, \mathbf {V}_{m},\overline{r}^c_m)} \    \sum\limits_{{\substack{m=1}}}^M \Big(\overline{r}^c_m + \tilde{R}^p_m\Big), \label{23a}  \\
		s.t.\ \ & \vartheta_m(1-\overline{D}_m) \leq \Lambda(\hat{E}_{m})- \overline{C}_m \overline{D}_m, \label{23b} \\
		\ \ & \eqref{22b} -\eqref{22f}. \label{23c} 
	\end{align}
\end{subequations}

As \textbf{(P3)} is highly non-convex because of the non-convex nature of its objective function and constraints. Thus, we introduce a slack variable $\varphi_m^p$, defined as follows:
\begin{subequations}
	\begin{align}
     \overline{r}^c_m + \varphi_m^p \geq {R}_{min} 
 ,\label{24a} 
	\end{align}
	\begin{align}
	\log_{2}\Biggl(1+\frac{ \beta_m \mathrm{tr} \Big(\big(\mathbf H_{m} - \delta_m^2 \mathbf I\big)\mathbf V_m\Big) }  {\beta_m  \sum\limits_{{\substack{i=1 \\ i\neq m}}}^{M} \mathrm{tr} \Big(\big(\mathbf H_{m} + \delta_m^2 \mathbf I\big) \mathbf{V}_{i} \Big) + \overline{\sigma}_{m}^{2}} \Biggr) \geq \varphi_m^p . \label{24b} 
	\end{align}
\end{subequations}

To track the convexity of \eqref{24b}, we can write as
\begin{align}
	&\log_{2}\Big( \beta_m \mathrm{tr} \Big(\big(\mathbf H_{m} - \delta_m^2 \mathbf I\big)\mathbf V_m\Big)+ \beta_m  \sum\limits_{{\substack{i=1 \\ i\neq m}}}^{M} \mathrm{tr} \Big(\big(\mathbf H_{m} + \nonumber \\
	& \delta_m^2 \mathbf I\big) \mathbf{V}_{i} \Big) + \overline{\sigma}_{m}^{2} \Big) -\log_{2}\Big(\beta_m  \sum\limits_{{\substack{i=1 \\ i\neq m}}}^{M} \mathrm{tr} \Big(\big(\mathbf H_{m} + \delta_m^2 \mathbf I\big) \mathbf{V}_{i} \Big) \nonumber \\
	& + \overline{\sigma}_{m}^{2} \Big) \geq \varphi_m^p .  \label{25}
\end{align}

Since Eq.~\eqref{25} is still non-convex due to the second term. Thus, we exploit first-order Taylor approximation (FTA) as follows
\begin{align}
	&\log_{2}\biggl( \beta_m \mathrm{tr} \Big(\big(\mathbf H_{m} - \delta_m^2 \mathbf I\big)\mathbf V_m\Big)+ \beta_m  \sum\limits_{{\substack{i=1 \\ i\neq m}}}^{M} \mathrm{tr} \Big(\big(\mathbf H_{m} + \nonumber \\
	& \delta_m^2 \mathbf I\big) \mathbf{V}_{i} \Big) + \overline{\sigma}_{m}^{2} \biggr) - \Delta_1 \big(\mathbf V_i\big) \geq \varphi_m^p,  \label{26}
\end{align}
where $\Delta_1 \big(\mathbf V_i\big)$ denotes the FTA given as follows 
\begin{align}
	\Delta_1 \big(\mathbf V_i\big)&=\log_{2}\Big(\beta_m  \sum\limits_{{\substack{i=1 \\ i\neq m}}}^{M} \mathrm{tr} \Big(\big(\mathbf H_{m} + \delta_m^2 \mathbf I\big) \mathbf{V}^{(r)}_{i} \Big) + \overline{\sigma}_{m}^{2} \Big)  \nonumber \\
		&+\sum\limits_{{\substack{i=1 \\ i\neq m}}}^{M} \mathrm{tr}\Bigg[ \left(  \frac{\beta_m (\mathbf H_m + \delta_m^2 \mathbf I)}{\ln2\beta_m \mathrm{tr} (\mathbf H_m + \delta_m^2 \mathbf I)\mathbf{V}^{(r)}_{i}+\overline{\sigma}_{m}^{2}}         \right)^H \nonumber \\
		&\times \big(\mathbf V_i - \mathbf V^{(r)}_i\big) \Bigg],  \label{27}
\end{align}
where $\mathbf V^{(r)}_i$ denotes the value of $\mathbf V_i$ in the $r^{th}$ iteration.

Accordingly, to track the convexity of \eqref{22c}, we can write as
\begin{align}
	&\log_{2}\Big( \beta_m \mathrm{tr} \Big(\big(\mathbf H_{m} - \delta_m^2 \mathbf I\big)\mathbf V_0\Big) + \beta_m  \sum\limits_{{\substack{i=1}}}^{M} \mathrm{tr} \Big(\big(\mathbf H_{m} + \delta_m^2 \mathbf I\big)\mathbf V_i\Big) \nonumber \\
	&+ \overline{\sigma}_{m}^{2}\Big) -\log_{2}\Big(\beta_m  \sum\limits_{{\substack{i=1}}}^{M} \mathrm{tr} \Big(\big(\mathbf H_{m} + \delta_m^2 \mathbf I\big)\mathbf V_i\Big) + \overline{\sigma}_{m}^{2} \Big) \nonumber \\
	&\geq  \sum\limits_{{\substack{m \in \mathcal M}}} \bar{r}_m^c .  \label{28}
\end{align}

It can be observed that \eqref{28} is non-convex. Thus, we employ successive convex approximation as follows
\begin{align}
	&\log_{2}\Big( \beta_m \mathrm{tr} \Big(\big(\mathbf H_{m} - \delta_m^2 \mathbf I\big)\mathbf V_0\Big) +  \beta_m  \sum\limits_{{\substack{i=1}}}^{M} \mathrm{tr} \Big(\big(\mathbf H_{m} + \delta_m^2 \mathbf I\big)\mathbf V_i\Big)  \nonumber \\
	& + \overline{\sigma}_{m}^{2}\Big) - \Delta_2 (\mathbf V_i) \geq  \sum\limits_{{\substack{m \in \mathcal M}}} \bar{r}_m^c ,  \label{29}
\end{align}
where $\Delta_2 (\mathbf V_i)$ denotes the FTA given as 
\begin{align}
	&\Delta_2 \big(\mathbf V_i\big)=\log_{2}\Biggl(\beta_m  \sum\limits_{{\substack{i=1}}}^{M} \mathrm{tr} \Big(\big(\mathbf H_{m} + \delta_m^2 \mathbf I\big)\mathbf V^{(r)}_i\Big) + \overline{\sigma}_{m}^{2} \Biggr) +  \nonumber \\
	&\sum\limits_{{\substack{i=1}}}^{M} \mathrm{tr}\Biggl[ \Biggl(  \frac{\beta_m (\mathbf H_m + \delta_m^2 \mathbf I)}{\ln2\beta_m \mathrm{tr} (\mathbf H_m + \delta_m^2 \mathbf I)\mathbf{V}^{(r)}_{i}+\overline{\sigma}_{m}^{2}}         \Biggr)^H\times \big(\mathbf V_i - \mathbf V^{(r)}_i\big) \Biggr].  \label{30}
\end{align}

Finally, the constraint defined by Eq.~\eqref{23b} can be expressed as
\begin{align}
	(1 - \beta_m) \Bigl( \sum_{i=0}^{M} \mathrm{tr}\bigl((\mathbf{H}_{m} - \delta_m^2 \mathbf I) \mathbf{V}_{i}\bigr) + \tilde{\sigma}_{m}^{2}\Bigr)\geq \Psi_m, \label{31}
\end{align}
where
\begin{align}
	\Psi_m = \overline{d}_m-\frac{1}{\overline{c}_m}\ln \left( \frac{\overline{C}_m}{\overline{C}_m \overline{D}_m+(1-\overline{D}_m)\vartheta_m}-1 \right). \label{32}
\end{align}

Consequently, the optimization problem \textbf{(P3)} can be restated as follows
\begin{subequations}\label{P4}
	\begin{align}
		\text{\textbf{(P4)}}	& \mathop {\max }\limits_{( \mathbf {V}_{0}, \mathbf {V}_{m},\overline{r}^c_m, \varphi_m^p)} \    \sum\limits_{{\substack{m=1}}}^M \Big(\overline{r}^c_m + \varphi_m^p\Big), \label{33a}  \\
		s.t.\ \ & \overline{r}^c_m + \varphi_m^p \geq {R}_{min} , \label{33b} \\
		\ \ & \eqref{22d}, \eqref{26}, \eqref{29}, \eqref{31}, \label{33c}\\
		\ \ & \mathbf V_m \succeq 0, \ \mathbf V_0 \succeq 0, \ \forall m\in \mathcal M, \label{33d}\\ 
		\ \ & \operatorname{rank}(\mathbf V_m)=1, \ \operatorname{rank}(\mathbf V_0)=1,  \forall m\in \mathcal M. \label{33e}
	\end{align}
\end{subequations}

It is important to note that the optimization problem \textbf{(P4)} remains non-convex due to the rank-1 constraints specified in \eqref{33e}. To address this, we relax the rank-1 constraints using the semi-definite relaxation (SDR) method. Consequently, the relaxed problem \textbf{(P4)} becomes a convex SDP problem, which can be efficiently solved using the CVX toolbox with the MOSEK solver. Moreover, when both matrices ${\mathbf{V}}^{*}_{0}$ and ${\mathbf{V}}_{m}$ are of rank one, the precoding vectors $\mathbf{v}^*_{0}$ and $\mathbf{v}^*_{m}$ can be computed by leveraging singular value decomposition (SVD). In contrast, if ${\mathbf{V}}^{*}_{0}$ and ${\mathbf{V}}_{m}$ exhibit higher rank, the Gaussian randomization technique can be employed \cite{ni2021resource}. 

\subsection{Optimization of $\beta_m:$ Stage 2}
Considering the transmit beamforming vectors $\mathbf{v}_0$ and $\mathbf{v}_m$ determined in Stage 1 and a fixed value of scattering matrix $\mathbf \Theta$, the corresponding optimization problem for calculating the power-splitting ratio is given as
\begin{subequations}\label{P5}
	\begin{align}
		\text{\textbf{(P5)}} & \mathop {\max }\limits_{( \mathbf \overline{r}^c_m, \beta_m)} \    \sum\limits_{{\substack{m=1}}}^M \Big(\overline{r}^c_m + \tilde{R}^p_m\Big), \label{222a}   \\
		s.t.\ \ & \eqref{22b},  \eqref{22c}, \eqref{31},  \label{222b} \\
		\ \ & 0 \leq\beta_m \leq 1, \forall m\in \mathcal M, \label{222c} 
	\end{align}
\end{subequations}

It is worth mentioning that problem \textbf{(P5)} is inherently convex in nature, which enables it to be efficiently solved using the CVX toolbox.

%%%%%%%%%%%%%%%%%%%%%%%%%%%%%%%%%%%%%%%%%%%%%%%%%%%%%%%%%%%%%%%%%%%%%%%%%%%%%%%%%%%%%%%%%%%%%%%%%%%%%%%%
\subsection{Optimization of $\mathbf{\Theta}:$ Stage 3}
In this stage, we leverage manifold optimization to determine the scattering matrix $\mathbf{\Theta}$ of the BD-RIS. Based on the transmit precoding vectors obtained from Stage 1 and the power-splitting ratio determined in Stage 2, the passive beamforming problem is formulated as follows:
\begin{subequations}\label{P6}
	\begin{align}
		\text{\textbf{(P6)}}	& \mathop {\max }\limits_{(\mathbf \Theta, \overline{r}^c_m)} \    \sum\limits_{{\substack{m=1}}}^M \Big(\overline{r}^c_m + \tilde{R}^p_m\Big), \label{34a}  \\
		s.t.\ \ & \eqref{22b},  \eqref{22c}, \eqref{31}, , \label{34b}\\
		\ \ & \mathbf \Theta \mathbf \Theta^H= \mathbf{I}_{L}. \label{34c}  
	\end{align}
\end{subequations}

Since problem \textbf{(P6)} is non-convex. To track the convexity of its objective function, let us define a set of slack variables  $\boldsymbol{\eta}_m=[\eta^p_{m,1}, \eta^p_{m,2}]^T$ as follows 
\begin{align}
	&\log_{2}\Bigg(\beta_m \mathrm{tr} \Big(\big(\mathbf H_{m} - \delta_m^2 \mathbf I\big)\mathbf V_m\Big)+ \nonumber \\
	&\beta_m  \sum\limits_{{\substack{i=1 \\ i\neq m}}}^{M} \mathrm{tr} \Big(\big(\mathbf H_{m} + \delta_m^2 \mathbf I\big) \mathbf{V}_{i} \Big) + \overline{\sigma}_{m}^{2} \Bigg)  \geq \eta^p_{m,1} , \label{35}
\end{align}
\begin{align}
	&\log_{2}\Bigg(\beta_m  \sum\limits_{{\substack{i=1 \\ i\neq m}}}^{M} \mathrm{tr} \Big(\big(\mathbf H_{m} + \delta_m^2 \mathbf I\big) \mathbf{V}_{i} \Big) + \overline{\sigma}_{m}^{2} \Bigg) \leq \eta^p_{m,2} , \label{36}
\end{align}

To track the convexity of \eqref{35} and \eqref{36}, we define a set of slack variables $\boldsymbol{\chi}_m=[\chi_{m,1}, \chi_{m,2}]^T$, 
\begin{align}
	& \beta_m \mathrm{tr} \Big(\big(\mathbf H_{m} - \delta_m^2 \mathbf I\big)\mathbf V_m\Big)+ \beta_m  \sum\limits_{{\substack{i=1 \\ i\neq m}}}^{M} \mathrm{tr} \Big(\big(\mathbf H_{m} + \delta_m^2 \mathbf I\big) \mathbf{V}_{i} \Big) \nonumber \\
	& + \overline{\sigma}_{m}^{2} \geq \chi_{m,1} , \label{37}
\end{align}
\begin{align}
	& \beta_m  \sum\limits_{{\substack{i=1 \\ i\neq m}}}^{M} \mathrm{tr} \Big(\big(\mathbf H_{m} + \delta_m^2 \mathbf I\big) \mathbf{V}_{i} \Big) + \overline{\sigma}_{m}^{2} \leq \chi_{m,2} , \label{38}
\end{align}
\begin{align}
	&\log_{2}\left( \chi_{m,1}\right) \geq \eta^p_{m,1} , \label{39}
\end{align}
\begin{align}
	&\log_{2}\left( \chi_{m,2}\right) \leq \eta^p_{m,2} , \label{40}
\end{align}

Further, to track the convexity of \eqref{40}, we employ successive convex approximation as follows
  \begin{align}
	\Biggl(\log_{2}\big(\chi_{m,2}^{(r)}\big) +\frac{\chi_{m,2}-\chi_{m,2}^{(r)}}{\ln (2) \chi_{m,2}^{(r)}} \Biggr) \leq \eta^p_{m,2}.\label{41} 
\end{align}

To track the convexity of \eqref{22c}, we define slack variables  $\boldsymbol{\eta}_c=[\eta^c_{m,1}, \eta^c_{m,2}]^T$ as follows
\begin{align}
	&\eta^c_{m,1} - \eta^c_{m,2}\geq  \sum\limits_{{\substack{m \in \mathcal M}}} \bar{r}_m^c ,  \label{42}
\end{align}
\begin{align}
	&\log_{2}\Bigg( \beta_m \mathrm{tr} \Big(\big(\mathbf H_{m} - \delta_m^2 \mathbf I\big)\mathbf V_0\Big) + \nonumber \\
	&\beta_m  \sum\limits_{{\substack{i=1}}}^{M} \mathrm{tr} \Big(\big(\mathbf H_{m} + \delta_m^2 \mathbf I\big)\mathbf V_i\Big) + \overline{\sigma}_{m}^{2}\Bigg)  \geq \eta^c_{m,1}, \label{43}
\end{align}
\begin{align}
	&\log_{2}\Bigg(\beta_m  \sum\limits_{{\substack{i=1}}}^{M} \mathrm{tr} \Big(\big(\mathbf H_{m} + \delta_m^2 \mathbf I\big)\mathbf V_i\Big) + \overline{\sigma}_{m}^{2} \Bigg) \leq \eta^c_{m,2} , \label{44}
\end{align}

Further, to track the convexity of \eqref{43} and \eqref{44}, we define slack variables $\boldsymbol{\hat{\Pi}_m}=[\hat{\Pi}_{m,1}, \hat{\Pi}_{m,2}]^T$, 
\begin{align}
	& \beta_m \mathrm{tr} \Big(\big(\mathbf H_{m} - \delta_m^2 \mathbf I\big)\mathbf V_0\Big) + \beta_m  \sum\limits_{{\substack{i=1}}}^{M} \mathrm{tr} \Big(\big(\mathbf H_{m} + \delta_m^2 \mathbf I\big)\mathbf V_i\Big) + \overline{\sigma}_{m}^{2} \nonumber \\
	& \geq \hat{\Pi}_{m,1} , \label{45}
\end{align}
\begin{align}
	& \beta_m  \sum\limits_{{\substack{i=1}}}^{M} \mathrm{tr} \Big(\big(\mathbf H_{m} + \delta_m^2 \mathbf I\big)\mathbf V_i\Big) + \overline{\sigma}_{m}^{2} \leq \hat{\Pi}_{m,2} , \label{46}
\end{align}
\begin{align}
	&\log_{2}\left(\hat{\Pi}_{m,1}\right) \geq \eta^c_{m,1} , \label{47}
\end{align}
\begin{align}
	&\log_{2}\left(\hat{\Pi}_{m,2}\right) \leq \eta^c_{m,2} , \label{48}
\end{align}

Moreover, to track the convexity of \eqref{48}, we FTA as follows
  \begin{align}
	\Biggl(\log_{2}\big(\hat{\Pi}_{m,2}^{(r)}\big) +\frac{\hat{\Pi}_{m,2}-\hat{\Pi}_{m,2}^{(r)}}{\ln (2) \hat{\Pi}_{m,2}^{(r)}} \Biggr) \leq \eta^c_{m,2}.\label{49} 
\end{align}

As a result, the optimization problem \textbf{(P6)} can be reformulated in the following manner:
\begin{subequations}\label{P7}
	\begin{align}
		\text{\textbf{(P7)}}	& \mathop {\max }\limits_{(\mathbf \Theta, \overline{r}^c_m, \boldsymbol{\eta}_m, \boldsymbol{\chi}_m, \boldsymbol{\eta}_c, \boldsymbol{\hat{\Pi}_m})} \    \sum\limits_{{\substack{m=1}}}^M \overline{r}^c_m + \Big(\eta^p_{m,1} - \eta^p_{m,2} \Big), \label{50a}  \\
		s.t. \ \ & \eqref{37}, \eqref{38}, \eqref{39}, \eqref{41}, \eqref{45}, \eqref{46}, \eqref{47}, \eqref{49}, \label{50b}\\
		\ \ & \overline{r}^c_m + \Big(\eta^p_{m,1} - \eta^p_{m,2} \Big)  \geq  {R}_{min}, \forall m\in \mathcal M, \label{50c} \\
		\ \ & \eta^c_{m,1} - \eta^c_{m,2}\geq  \sum\limits_{{\substack{m \in \mathcal M}}} \bar{r}_m^c, \forall m\in \mathcal M, \label{50d}\\
		\ \ & \mathbf \Theta \mathbf \Theta^H= \mathbf{I}_{L} \label{50e}.    
	\end{align}
\end{subequations}

\begin{figure*}
	\begin{align} \label{51} 
		\mathcal{L}(\mathbf \Theta) = &  \sum_{m=1}^M \Biggl[
		\Big(- \overline{r}_m^c - \eta_{m,1} + \eta_{m,2}\Big) - \rho_m \Bigl( 
		(1 - \beta_m) \Bigl( \sum_{i=0}^{M} \mathrm{tr}\Bigl((\mathbf{H}_{m} - \delta_m^2 \mathbf I) \mathbf{V}_{i}\Bigr) + \tilde{\sigma}_{m}^{2} \Bigr) - \Psi_m 
		\Bigr)
		+ \lambda_{m,1} \Bigl\{ 
		\chi_{m,1} 
		\nonumber\\
		&- \beta_m  \mathrm{tr} \Big(\Big(\mathbf H_{m} - \delta_m^2 \mathbf I\Big)\mathbf V_m\Big)+ \beta_m  \sum\limits_{{\substack{i=1 \\ i\neq m}}}^{M} \mathrm{tr} \Big(\big(\mathbf H_{m} + \delta_m^2 \mathbf I\Big) \mathbf{V}_{i} \Big) + \overline{\sigma}_{m}^{2} 
		\Bigr\} + \lambda_{m,2} \Bigl\{
		\beta_m  \sum\limits_{{\substack{i=1 \\ i\neq m}}}^{M} \mathrm{tr} \Big(\Big(\mathbf H_{m}  + \delta_m^2 \mathbf I\Big) \nonumber\\
		& \mathbf{V}_{i} \Big) + \overline{\sigma}_{m}^{2} - \chi_{m,2}
		\Bigr\} + \lambda_{m,3} \Bigl\{ \eta^p_{m,1} - \log_2(\chi_{m,1}) \Bigr\} + \lambda_{m,4} \Bigl\{ 
		\log_2(\chi^{(r)}_{m,2}) + \frac{\chi_{m,2} - \chi^{(r)}_{m,2}}{\ln 2 \ \chi^{(r)}_{m,2} } - \eta^{(p)}_{m,2}
		\Bigr\}  - \lambda_{m,5} \nonumber\\
		&\Bigl\{ \overline{r}^c_m + \Big(\eta^p_{m,1} - \eta^p_{m,2} \Big) 
		- {R}_{min}
		\Bigr\} - \mu_{m,1} \Bigl\{ \Big(\eta^c_{m,1} - \eta^c_{m,2} \Big) - \sum\limits_{{\substack{m \in \mathcal M}}} \overline{r}_m^c \Bigr\} + \mu_{m,2} \Bigl\{ 
		\Pi_{m,1} - \beta_m \mathrm{tr} \Big(\big(\mathbf H_{m} - \delta_m^2 \mathbf I\big)\nonumber\\
		& \mathbf V_0\Big) + \beta_m  \sum\limits_{{\substack{i=1}}}^{M} \mathrm{tr} \Big(\big(\mathbf H_{m} + \delta_m^2 \mathbf I\big)\mathbf V_i\Big) 
		+ \overline{\sigma}_{m}^{2}
		\Bigr\} + \mu_{m,3} \Bigl\{ \beta_m  \sum\limits_{{\substack{i=1}}}^{M} \mathrm{tr} \Big(\big(\mathbf H_{m} + \delta_m^2 \mathbf I\big)\mathbf V_i\Big) + \overline{\sigma}_{m}^{2} - \Pi_{m,2}
		\Bigr\} \nonumber\\
		&   + \nu_{m,1} \Bigl\{ \eta^c_{m,1}  - \log_{2}\left( \Pi_{m,1}\right) 
		\Bigr\} +  \nu_{m,2} \Bigl\{ \Big( 
		\log_2(\Pi^{(r)}_{m,2}) + \frac{\Pi_{m,2} - \Pi^{(r)}_{m,2}}{\ln 2 \Pi^{(r)}_{m,2}}\Big) - \tilde{\eta}_{m,2}
		\Bigr\}\Biggr].
	\end{align}\hrulefill 
\end{figure*}

It is important to highlight that the optimization problem \textbf{(P7)} is still challenging to solve due to non-convexity arises from the unitary constraint, defined in \eqref{50e}, which forms a $L^2$-dimensional complex Stiefel manifold \cite{boumal2014manopt}. Specifically, a manifold can be mathematically described as a topological space that is locally equivalent to Euclidean space through homeomorphisms. For every point on the manifold, the tangent vectors represent the instantaneous possible directions of movement, and the set of these vectors defines the tangent space. This tangent space is Euclidean and contains the Riemannian gradient \cite{absil2008optimization, zhu2017riemannian}, which directs toward the steepest decline of the objective function. Herein, we adopt a manifold optimization approach to efficiently solve problem \textbf{(P7)}. The main steps are summarized as follows: 1) we construct the set of all possible solutions with respect to $\mathbf{\Theta}$ as a manifold; 2) we subsequently proceed by mapping the original constrained optimization setup from Euclidean space onto a manifold, resulting in an unconstrained problem; and 3) we apply the Conjugate-Gradient (CG) method for iterative updates\cite{zhu2017riemannian}, where the computation of the Riemannian gradient entails projecting the standard Euclidean gradient onto the tangent space associated with the manifold.

Next, to compute the Euclidean gradient for problem \textbf{(P7)}, we first reformulate it into an unconstrained form by constructing its Lagrangian function $\mathcal{L}(\mathbf \Theta)$ defined by Eq.~\eqref{51}. Accordingly, the Euclidean gradient can be computed as: 
\begin{align}
	\mathcal{G} \mathcal L (\mathbf \Theta)=&\sum_{m=1}^M \sum_{j=1}^5 a_{m,j} 
	\Bigl(\mathbf{h}_{rm} \mathbf{f}_{bm}^H \mathbf{S}_{m,j} \mathbf{G}_{br}^H
	+ \mathbf{G}_{br} \mathbf{S}_{m,j} \mathbf{f}_{bm} \mathbf{h}_{rm}^H
	\nonumber \\
	&+ 2\mathbf{G}_{br} \mathbf{S}_{m,j} \mathbf{G}_{br}^H \mathbf{\Theta} \mathbf{h}_{rm} \mathbf{h}_{rm}^H\Bigr),  \label{52} 
\end{align}
where,
\[
\begin{cases}
	a_{m,1} = -\rho_m (1-\beta_m), \quad
	\mathbf{S}_{m,1} = \sum_{i=0}^M \mathbf{V}_i \\
	a_{m,2} = \lambda_{m,1} \beta_m, \quad
	\mathbf{S}_{m,2} = \sum_{\substack{i=1 \\ i \neq m}}^{M} \mathbf{V}_i - \mathbf{V}_m \\
	a_{m,3} = \lambda_{m,2} \beta_m, \quad
	\mathbf{S}_{m,3} = \sum_{\substack{i=1 \\ i \neq m}}^{M} \mathbf{V}_i \\
	a_{m,4} = \mu_{m,2} \beta_m, \quad
	\mathbf{S}_{m,4} = \sum_{i=1}^M \mathbf{V}_i - \mathbf{V}_0 \\
	a_{m,5} = \mu_{m,3} \beta_m, \quad
	\mathbf{S}_{m,5} = \sum_{i=1}^M \mathbf{V}_i
\end{cases}
\]
Additionally, the values of $\overline{r}_m^c$, and Lagrange dual variables $\rho_m$, $\boldsymbol{\lambda}_m = \{ \lambda_{m,1}, \lambda_{m,2}, \lambda_{m,3}, \lambda_{m,4}, \lambda_{m,5}\}$, $\boldsymbol{\mu}_m = \{ \mu_{m,1}, \mu_{m,2}, \mu_{m,3}\}$, and $\boldsymbol{\nu}_m = \{ \nu_{m,1}, \nu_{m,2}\}$ are updated leveraging Sub-gradient method \cite{asif2025noma}, as follows
\begin{align}
	&\overline{r}_m^c (r+1)= \Big[\overline{r}_m^c (r)+ \varrho_m   
		\Big(1 + \lambda_{m,5} - \sum_{m=1}^M \mu_{m,1}\Big)\Big]^+,\label{53}  
\end{align}
\begin{align}
	\rho_m &(r+1)=  \Big[\rho_m (r)+ \varrho_m   
	\Bigl( (1 - \beta_m) \Bigl( \sum_{i=0}^{M} \mathrm{tr}\Bigl((\mathbf{H}_{m} - \delta_m^2 \mathbf I)  \nonumber \\
	& \mathbf{V}_{i}\Bigr)+ \tilde{\sigma}_{m}^{2} \Bigr) - \Psi_m 
	\Bigr)\Big]^+,\label{54}   
\end{align}
\begin{align}
	\lambda_{m,1}&(r+1)= \Big[\lambda_{m,1}(r)+ \varrho_m   
	\Bigl\{ 
	\chi_{m,1} 
	- \beta_m \mathrm{tr} \Big(\Big(\mathbf H_{m} - \delta_m^2 \mathbf I\Big)\mathbf  \nonumber \\
	&  V_m\Big) + \beta_m  \sum\limits_{{\substack{i=1 \\ i\neq m}}}^{M} \mathrm{tr} \Big(\big(\mathbf H_{m} + \delta_m^2 \mathbf I\Big) \mathbf{V}_{i} \Big) + \overline{\sigma}_{m}^{2} 
	\Bigr\}\Big]^+,\label{55}   
\end{align}
\begin{align}
	\lambda_{m,2}&(r+1)= \Big[\lambda_{m,2}(r)+ \varrho_m \Bigl\{\beta_m  \sum\limits_{{\substack{i=1 \\ i\neq m}}}^{M} \mathrm{tr} \Big(\Big(\mathbf H_{m} + \delta_m^2 \mathbf I\Big) \nonumber \\
	& \mathbf{V}_{i} \Big) + \overline{\sigma}_{m}^{2} - \chi_{m,2}
		\Bigr\} \Big]^+,\label{56}   
\end{align}
\begin{align}
	\lambda_{m,3}&(r+1) =  \Big[\lambda_{m,3}(r)+ \varrho_m \Bigl\{ \eta^p_{m,1} - \log_2(\chi_{m,1}) \Bigr\} \Big]^+,\label{57}   
\end{align}
\begin{align}
	\lambda_{m,4}&(r+1)=  \Big[\lambda_{m,4}(r)+ \varrho_m \Bigl\{ 
		\log_2(\chi^{(r)}_{m,2}) + \frac{\chi_{m,2} - \chi^{(r)}_{m,2}}{\ln 2 \ \chi^{(r)}_{m,2} } \nonumber \\
		& - \eta^{(p)}_{m,2}
		\Bigr\} \Big]^+,\label{58}   
\end{align}
\begin{align}
	\lambda_{m,5}&(r+1)=  \Big[\lambda_{m,5}(r)+ \varrho_m \Bigl\{ \overline{r}^c_m + \Big(\eta^p_{m,1}- \eta^p_{m,2} \Big)  \nonumber \\
	&  - {R}_{min}\Bigr\} \Big]^+,\label{59}   
\end{align}
\begin{align}
	\mu_{m,1}(r+1)= & \Big[\mu_{m,1}(r) + \varrho_m \Bigl\{ \Big(\eta^c_{m,1} - \eta^c_{m,2} \Big) \nonumber \\
	&- \sum\limits_{{\substack{m \in \mathcal M}}} \bar{r}_m^c \Bigr\} \Big]^+,\label{60}   
\end{align}
\begin{align}
	\mu_{m,2}&(r+1)= \Big[\mu_{m,2}(r) + \varrho_m \Bigl\{\Pi_{m,1} - \beta_m \mathrm{tr} \Big(\big(\mathbf H_{m} - \delta_m^2 \mathbf I\big)\nonumber \\
	& \mathbf V_0\Big)  + \beta_m  \sum\limits_{{\substack{i=1}}}^{M} \mathrm{tr} \Big(\big(\mathbf H_{m} + \delta_m^2 \mathbf I\big)\mathbf V_i\Big) + \overline{\sigma}_{m}^{2}
		\Bigr\}  \Big]^+,\label{61}   
\end{align}
\begin{align}
	\mu_{m,3}&(r+1)= \Big[\mu_{m,3}(r) + \varrho_m \Bigl\{ \beta_m  \sum\limits_{{\substack{i=1}}}^{M} \mathrm{tr} \Big(\big(\mathbf H_{m} + \delta_m^2 \mathbf I\big)\nonumber \\
	&\mathbf V_i\Big) + \overline{\sigma}_{m}^{2} - \Pi_{m,2}
		\Bigr\} \Big]^+,\label{62}   
\end{align}
\begin{align}
	\nu_{m,1}&(r+1)= \Big[\nu_{m,1}(r) + \varrho_m \Bigl\{ \eta^c_{m,1} - \log_{2}\left( \Pi_{m,1}\right) 
		\Bigr\}  \Big]^+,\label{63}   
\end{align}
\begin{align}
	\nu_{m,2}&(r+1) = \Big[\nu_{m,2}(r) + \varrho_m \Bigl\{ \Big( 
		\log_2(\Pi^{(r)}_{m,2}) \nonumber \\
		& + \frac{\Pi_{m,2} - \Pi^{(r)}_{m,2}}{\ln 2 \Pi^{(r)}_{m,2}}\Big)   - \tilde{\eta}_{m,2}
		\Bigr\}  \Big]^+,\label{64}   
\end{align}
where $\overline{r}_m^c (r)$, $\rho_m (r)$, $\lambda_{m,j_1}(r), \forall j_1=1,2,..., 5$, $\mu_{m,j_2}(r), \forall \ \ j_2=1,2,3$, and $\nu_{m,j_3}(r), \forall \ j_3=1,2$, represent the values of dual variables in $r^{th}$ iteration, and $\varrho_m$ denotes the learning rate of sub-gradient method.

Next, we project the Euclidean gradient $\mathcal{G} \mathcal L (\mathbf \Theta)$ onto the tangent space to determine the Riemannian gradient $\mathcal{G}_{\mathcal{M}_p} \mathcal{L} (\mathbf{\Theta})$ as follows
\begin{align}
	\mathcal{G}_{\mathcal{M}_p} \mathcal{L}(\mathbf{\mathbf \Theta}) 
	&= \mathbf{\Pi}_{\mathbf{\Theta}}\bigl(\mathcal{G} \mathcal L (\mathbf \Theta)\bigr) \nonumber \\
	&= \mathcal{G} \mathcal L (\mathbf \Theta) 
	- \mathbf{\Theta} \cdot \frac{\mathbf{\Theta}^H \mathcal{G} \mathcal L (\mathbf \Theta) + \mathcal{G} \mathcal L (\mathbf \Theta)^H \mathbf{\Theta}}{2}, \label{65} 
\end{align}
where $\mathbf{\Pi}_{\mathbf{\Theta}}\bigl(\mathcal{G} \mathcal L (\mathbf \Theta)\bigr)$ represents the projection function. Further, we determine the descent direction at the $r^{\text{th}}$ iteration by employing the CG method as follows.
\begin{align}
	\boldsymbol{\Omega}^{r} 
	&= -\mathcal{G}_{\mathcal{M}_p} \mathcal{L}(\mathbf{\mathbf \Theta}^r) +\varepsilon^{r}\mathbf{\Pi}_{\mathbf{\Theta}^r}\bigl(\boldsymbol{\Omega}^{r-1} \bigr), \label{66} 
\end{align}
where $\varepsilon^{r}$ denotes the parameter for CG method which can be updated by exploiting the Riemannian extension of the Polak–Ribière method \cite{grippo1997globally}, as follows
\begin{align}
	&\varepsilon^{r}= \nonumber \\
	&\frac{\mathcal{R}\left\{\operatorname{tr}\left(\mathcal{G}_{\mathcal{M}_p} \mathcal{L}(\mathbf{\mathbf \Theta}^r)^H \left[\mathcal{G}_{\mathcal{M}_p} \mathcal{L}(\mathbf{\mathbf \Theta}^r)- \mathbf{\Pi}_{\mathbf{\Theta}} \left( \mathcal{G}_{\mathcal{M}_p} \mathcal{L}(\mathbf{\mathbf \Theta}^{r-1}) \right)              \right]\right)\right\}}{\operatorname{tr}\left(\mathcal{G}_{\mathcal{M}_p} \mathcal{L}(\mathbf{\mathbf \Theta}^{r-1})^H \mathcal{G}_{\mathcal{M}_p} \mathcal{L}(\mathbf{\mathbf \Theta}^{r})\right)} . \label{67} 
\end{align}

Finally, to map the updated point from the tangent space back onto the manifold, we apply the retraction operation as follows:
	\begin{align}
		\mathbf{\Theta}^{r+1} & =\operatorname{Ret}_{\mathbf{\Theta}}\left(\varepsilon^r \mathbf{\Omega}^r\right) \nonumber \\
		& =\frac{\mathbf{\Theta}^r + \phi^r \mathbf{\Omega}^r}{\sqrt{\mathbf{I}+\left(\varepsilon^r\right)^2\left(\mathbf{\Omega}^r\right)^H \mathbf{\Omega}^r}}, \label{68} 
	\end{align}
where $\phi^r$ is the step size, which can be updated at each iteration by employing a step size adjustment method like backtracking algorithm \cite{absil2008optimization}. 
\subsection{Proposed Algorithm and Complexity Analysis }
In \textbf{Algorithm~1}, we propose a three-stage resource allocation framework for the considered system. The computational complexity mainly arises from solving sub-problems \textbf{(P4)}, \textbf{(P5)}, and \textbf{(P7)}. Firstly, the SDP problem \textbf{(P4)} is addressed using the MOSEK-assisted CVX toolbox, resulting in a computational complexity of $\scalemath{0.85}{\mathcal{O}\bigl(\hat{r}_1 K^{3.5} \bigr)}$ \cite{wright1997primal}, 
where $\hat{r}_1$ denotes the number of iterations required for convergence. Secondly, problem \textbf{(P5)} is solved by leveraging the MOSEK-enabled CVX toolbox, yielding a computational complexity of 
$\scalemath{0.85}{\mathcal{O}\bigl(\hat{r}_2 M^{3} \bigr)}$\cite{wright1997primal}, 
where $\hat{r}_2$ indicates the iteration count for the convergence of \textbf{(P5)}. Thirdly, problem \textbf{(P7)} is solved using the Conjugate Gradient (CG) method, which entails a computational complexity of $\scalemath{0.85}{\mathcal{O}\bigl(\hat{r}_3 L^{3} \bigr)}$\cite{li2022beyond}, where $\hat{r}_3$ denotes the number of iterations required for convergence. Consequently, the overall computational complexity of \textbf{Algorithm~1} is expressed as $\scalemath{0.85}{\mathcal{O}\Big[\mathcal{\tilde{I}}_{max}\Big(\hat{r}_1 K^{3.5}  + \hat{r}_2 M^{3}  + \hat{r}_3 L^{3} \Big)\Big]}$, where $\mathcal{\tilde{I}}_{\max}$ denotes the number of iterations needed for the convergence of the overall alternating optimization framework.
\begin{algorithm}[t]
	\caption{Alternating Optimization Framework}
	\begin{algorithmic}[1]
		\State \textbf{Initialization:} Initialize the system parameters, $ \hat{\mathbf V}_0$, $\hat{\mathbf V}_m$, $\hat{\mathbf \Theta}$, $\hat{\rho}_m$, $\hat{\boldsymbol{\lambda}}_m$, $\hat{\boldsymbol{\mu}}_m$, $\hat{\boldsymbol{\nu}}_m$
		\While{not converged and $\mathcal{\tilde{I}}_{\max}$ not reached}
		
		\While{not converged and $\hat{r}_1$ not reached}
		\State Compute $\mathbf V_0$, $\mathbf V_m$, and $\overline{r}_m^c$ by solving \textbf{\textbf{(P4)}}
		\State Update $\hat{\mathbf V}_0 \leftarrow \mathbf V_0$ and $\hat{\mathbf V}_m \leftarrow \mathbf V_m$
		\EndWhile
		
		\While{not converged and $\hat{r}_2$ not reached}
		\State Compute $\beta_m$ by solving \textbf{(P5)}
		\EndWhile
		
		\While{not converged and $\hat{r}_3$ not reached}
		\State Compute Riemannian gradient $\mathcal{G}_{\mathcal{M}_p} \mathcal{L} (\mathbf{\Theta})$ by \eqref{65}
		\State Update descent direction $\boldsymbol{\Omega}$ by \eqref{66}
		\State Update CG parameter $\varepsilon$ based on \eqref{67}
		\State Update scattering matrix $\mathbf \Theta$ based on 
		\State retraction process in \eqref{68}
		\State Update $\phi$ based on backtracking algorithm []
		\State Update $\overline{r}_m^c$ and dual variables $\rho_m$, $\boldsymbol{\lambda}_m$, $\boldsymbol{\mu}_m$, $\boldsymbol{\nu}_m$ 
		\State based on \eqref{53}--\eqref{64}
		\State Update $\hat{\mathbf \Theta} \leftarrow \mathbf \Theta$
		\EndWhile
		
		\EndWhile
		\State \Return $\mathbf V_0$, $\mathbf V_m$, $\beta_m$, $\mathbf \Theta$ 
	\end{algorithmic}
\end{algorithm}
\begin{figure}[t]
	\centering
	\includegraphics [width=0.35\textwidth]{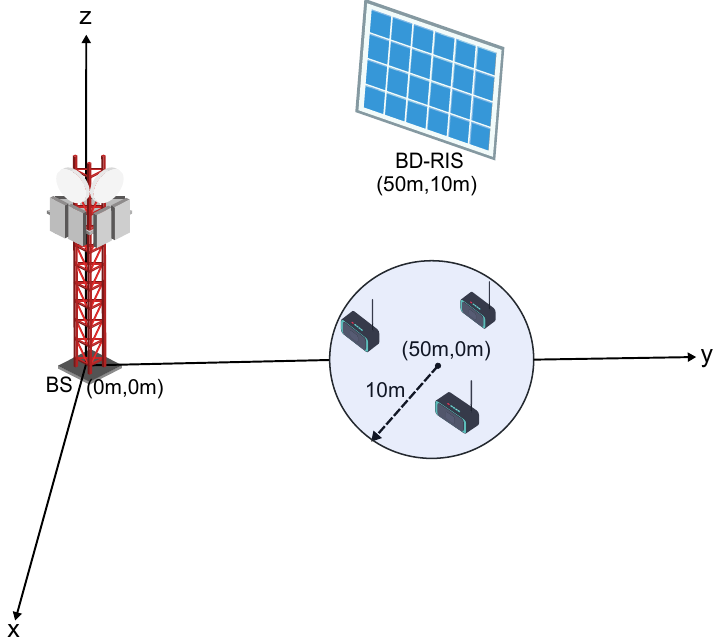}
	\caption{Simulation environment}
	\label{f2}
\end{figure} 
	\begin{figure}[t!]
	\centering
	\includegraphics [width=0.35\textwidth]{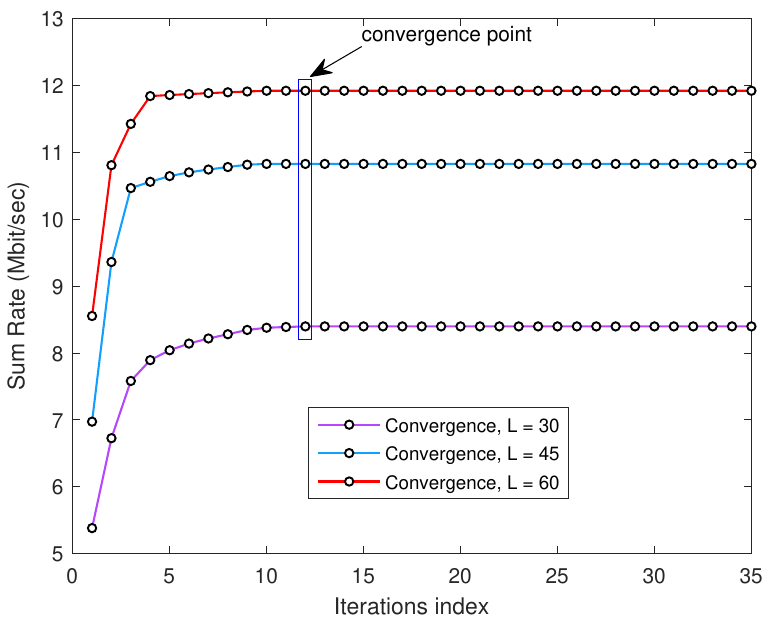}
	\caption{Convergence behavior for different values of $L$}
	\label{f3}
\end{figure}

\begin{figure}[t!]
	\centering
	\includegraphics [width=0.35\textwidth]{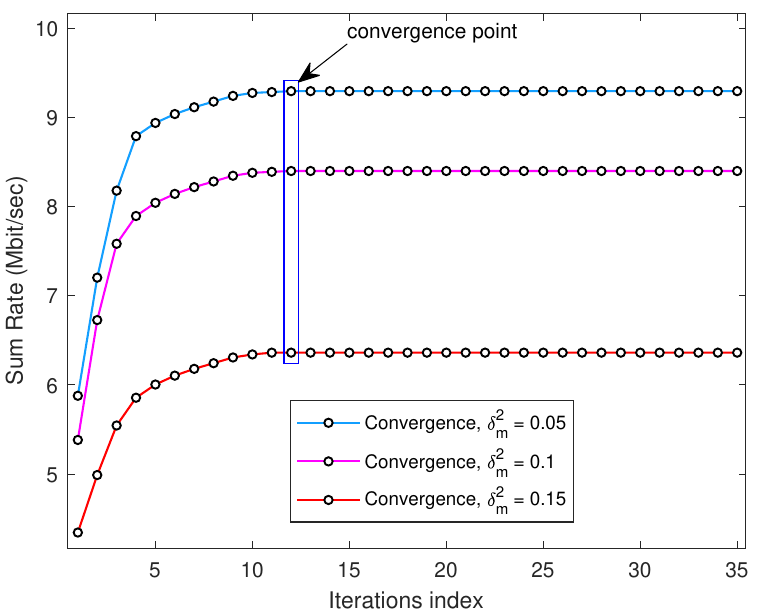}
	\caption{Convergence behavior for different values of $\delta_m^2$}
	\label{f4}
\end{figure}

\section{Performance Evaluation}

In this section, we employ numerical simulation setup to evaluate the performance of considered system under imperfect channel state information. As illustrated in Fig.~\ref{f2}, in current scenario, we set the coordinates of BS and BD-RIS as (0m, 0m) and (50m, 10m), respectively, where the users are randomly distributed within a circular region of radius 10m centered at (50m, 0m). Additionally, we consider that the channels $\mathbf{G}_{br}$, $\mathbf{h}_{rm}$, and $\mathbf{f}_{bm}$, $\forall m\in \mathcal M$ follow the Rician fading distribution model, with their respective large-scale path loss given by $10^{-3} d^{-2}$, $10^{-3} d^{-2.5}$, and $10^{-3} d^{-4}$ \cite{yang2021reconfigurable}, where $d$ represents distance in meters. Further, the NLOS components of the channels $\mathbf{G}_{br}$, $\mathbf{h}_{rm}$, and $\mathbf{f}_{bm}$ are modeled as independent and identically distributed complex Gaussian random variables\cite{asif2025noma}, whereas the LOS components are computed based on the steering vectors, which take into account the angle of departure and angle of arrival of the corresponding paths \cite{yang2021reconfigurable}. For the non-linear energy harvesting model, the parameters are chosen as $\overline{c}_m=6400$, $\overline{d}_m=0.003$, and $\overline{C}_m=0.2\,\mathrm{mW}$ \cite{camana2022rate, xiong2017rate}. Further, unless specified otherwise, the remaining simulation parameters are set as follows: $L = 30$, $\zeta_{br}=\zeta_{rm}=5$, $\zeta_{bm}=3$, $K = 3$, $\vartheta_m = -28$ dBm, $\tilde{\sigma}_m^2 = \tilde{\sigma}_{dec.}^2 = 10^{-8}$, Bandwidth = $1$ MHz, $R_{min} = 0.1$ Mbps, $\delta_m^2 =\tilde{\varrho} \left\|\mathbf{H}_m\right\|$, where $\tilde{\varrho} \in [0, 1)$  \cite{zheng2023zero}.

   \begin{figure}[t]
	\centering
	\includegraphics [width=0.35\textwidth]{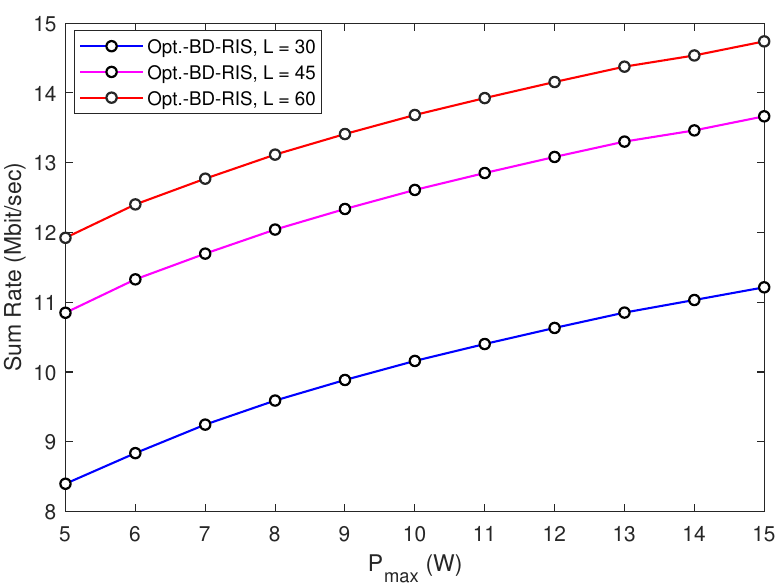}
	\caption{System's performance for different values of $L$}
	\label{f5}
\end{figure} 

\begin{figure}[t]
	\centering
	\includegraphics [width=0.35\textwidth]{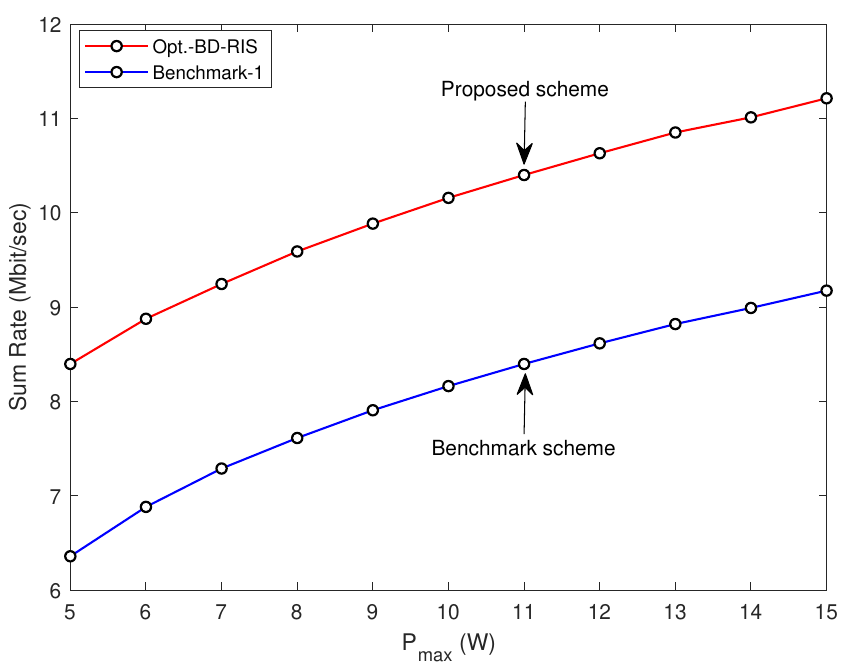}
	\caption{Performance comparison for increasing values of $P_{max}$}
	\label{f6}
\end{figure}

	\begin{figure}[t]
	\centering
	\includegraphics [width=0.35\textwidth]{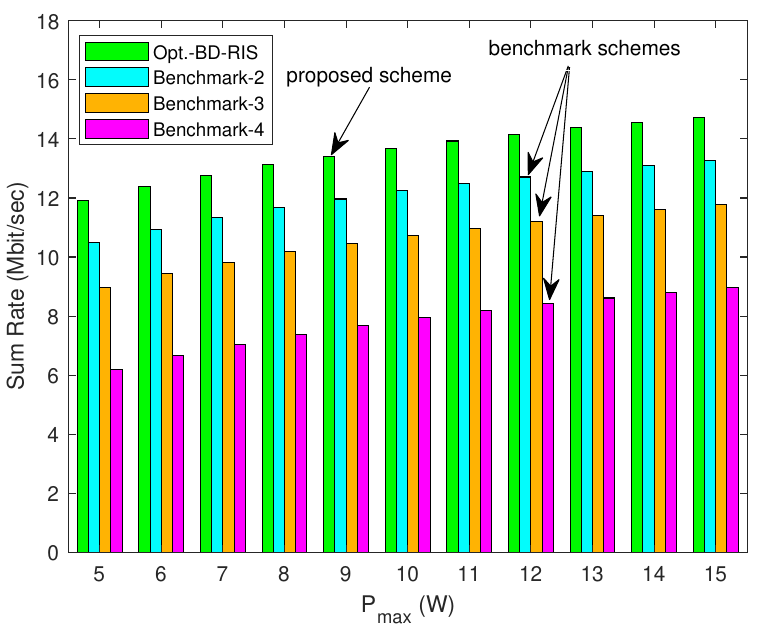}
	\caption{Performance comparison between the proposed scheme and benchmark counterparts}
	\label{f7}
\end{figure}

To evaluate the performance of the proposed technique, denoted as ``Opt.-BD-RIS,'' we introduce several benchmark schemes for comparison. Specifically, \textbf{Benchmark-1}: solves problems \textbf{(P4)}, \textbf{(P5)}, and \textbf{(P7)} under a single-connected RIS configuration. \textbf{Benchmark-2}: solves \textbf{(P4)} and \textbf{(P7)} with randomly assigned values for~$\beta_m$. \textbf{Benchmark-3}: addresses \textbf{(P7)}, where both~$\beta_m$ and the precoding vectors are randomly allocated. Finally, \textbf{Benchmark-4}: allocates all optimization variables randomly while maintaining the same system conditions as the proposed scheme.

	\begin{figure}[t]
	\centering
	\includegraphics [width=0.35\textwidth]{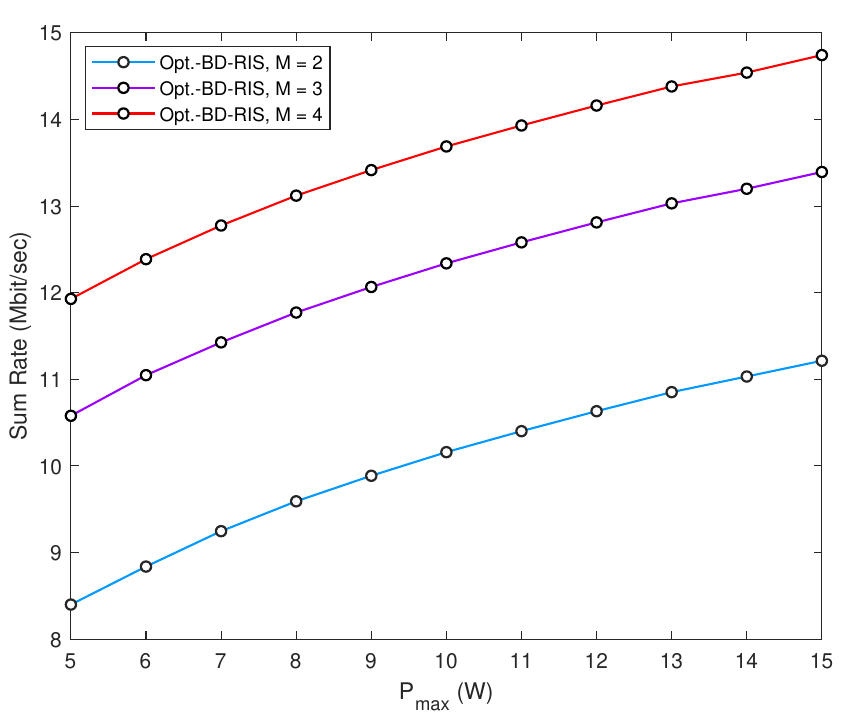}
	\caption{System's performance for increasing number of users}
	\label{f8}
\end{figure}

\begin{figure}[t]
	\centering
	\includegraphics [width=0.35\textwidth]{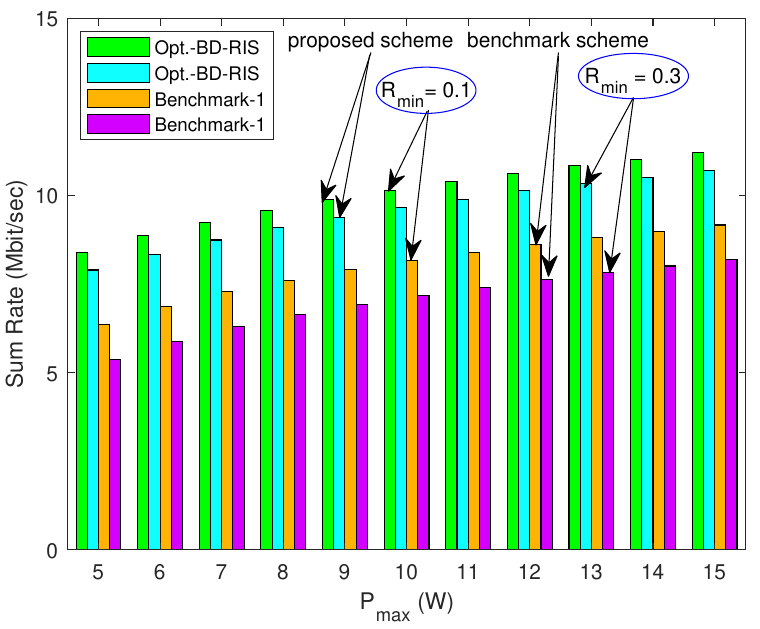}
	\caption{Performance comparison under different values of $R_{min}$}
	\label{f9}
\end{figure}

\begin{figure}[t]
	\centering
	\includegraphics [width=0.35\textwidth]{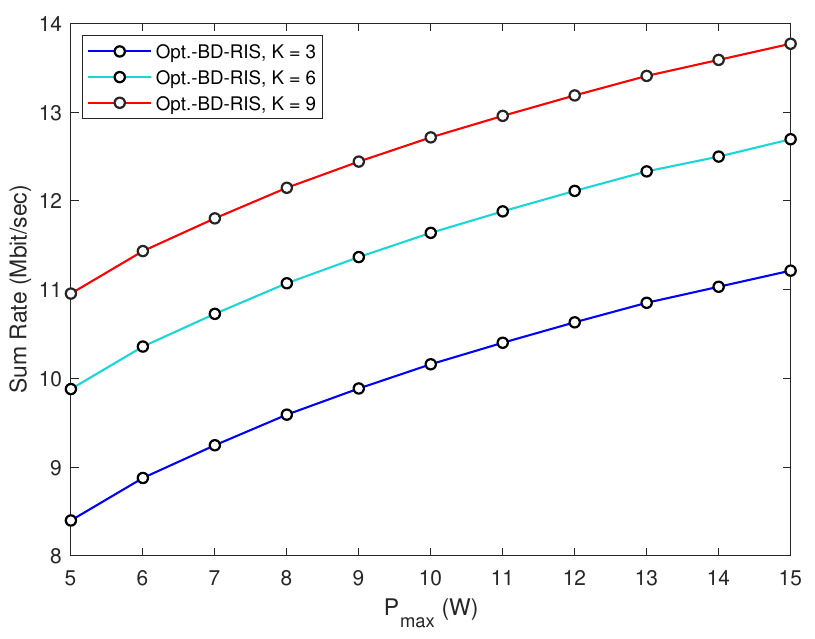}
	\caption{System's performance for increasing values of $K$}
	\label{f10}
\end{figure}

\begin{figure}[t]
	\centering
	\includegraphics [width=0.35\textwidth]{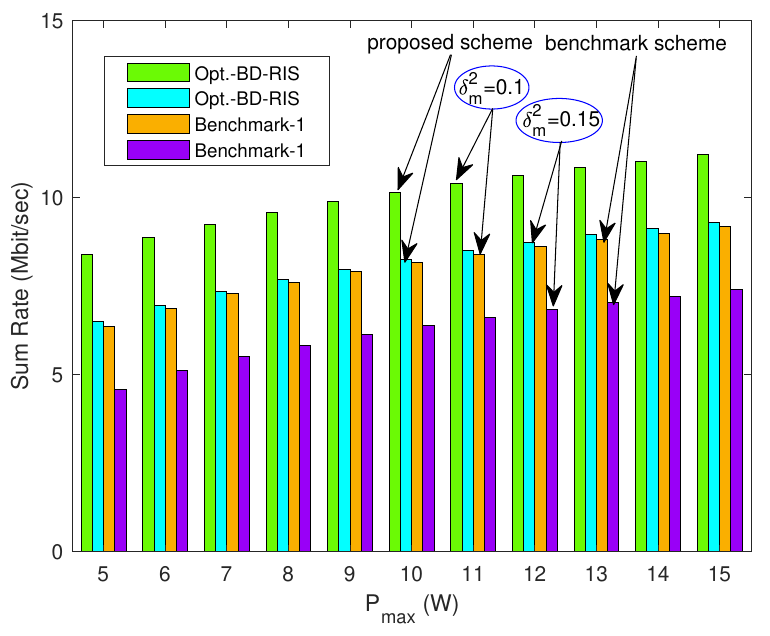}
	\caption{System's sum-rate for different values of $\delta_m^2$}
	\label{f11}
\end{figure}

The convergence behavior of the proposed approach is presented in Fig.~\ref{f3} for different numbers of BD-RIS elements. Simulation results show that the system’s sum-rate notably improves with larger values of $L$, since the beamforming capability of the BD-RIS is enhanced as $L$ increases. Likewise, Fig.~\ref{f4} illustrates the convergence performance of the proposed scheme under different values of $\delta_m^2$. It can be observed that the system’s performance declines as $\delta_m^2$ increases. This is due to the fact that larger channel estimation errors degrade the achievable rates of both common and private data streams, ultimately resulting in a lower overall sum-rate. Moreover, the numerical simulation results shown in Fig.~\ref{f3} and Fig.~\ref{f4} demonstrate that the proposed optimization framework achieves convergence within a reasonable number of iterations.

The effect of increasing the number of BD-RIS elements, $L$, is illustrated in Fig.~\ref{5} for various values of the available power budget $P_{max}$. As $L$ grows, a significant improvement in the system’s sum-rate performance is observed. This enhancement is primarily attributed to the BD-RIS's elements which are fully-connected through the reconfigurable impedance, where the beamforming capability of the BD-RIS is strengthened as $L$ increases.

Fig.~\ref{6} presents a comparative performance analysis of the proposed approach, Opt.-BD-RIS, against its benchmark scheme as the available power budget, $P_{max}$, increases. The simulation results reveal that the system’s sum rate improves with larger values of $P_{max}$, as the sum rate is a monotonically increasing function of the power budget. Consequently, a higher $P_{max}$ provides more power for transmission, thereby enhancing the system’s sum-rate performance. Moreover, Fig.~\ref{6} shows that the proposed Opt.-BD-RIS consistently outperforms the benchmark single-connected RIS scheme across all values of $P_{max}$. This performance gain stems from the fully-connected reconfigurable elements in BD-RIS, which enable more flexible and efficient beamforming by dynamically adjusting the connecting impedance among the elements. Additionally, Fig.~\ref{7} presents a performance comparison of the proposed approach, Opt.-BD-RIS, with three benchmark schemes, over increasing values of $P_{max}$. It is evident from the simulation outcomes that the proposed strategy consistently attains higher performance levels than all benchmark schemes across the entire range of $P_{max}$.

The impact of increasing values of $P_{max}$ on the system performance for different numbers of users is illustrated in Fig.~\ref{8}. It can be observed that the system’s sum-rate increases with the number of users in the network. Since the base station is equipped with multiple antennas, adding more users enhances the multi-user diversity gain. Due to temporal and spatial variations in wireless channels, a larger user pool increases the likelihood of finding users with favorable instantaneous channel conditions. Consequently, the base station can opportunistically allocate more resources to these users while assigning just enough resources to others to satisfy their QoS requirements. However, as the number of users continues to grow, the incremental sum-rate gain gradually diminishes because the highest achievable channel quality approaches a saturation point.

Further, Fig.~\ref{f9} illustrates the sum-rate performance of the considered system for different values of $R_{min}$ over a range of $P_{max}$. The results show that the system’s sum-rate decreases significantly as $R_{min}$ increases. This reduction occurs because the resource allocation framework must allocate more resources to users with poor channel conditions in order to meet the stricter QoS requirements imposed by higher values of $R_{min}$. As a result, fewer resources are left for other users with good channel conditions, which ultimately lowers the overall sum-rate. Additionally, it can be observed that the proposed technique consistently outperforms the benchmark scheme across all values of $P_{max}$. Moerover, the impact of increasing the number of BS antennas, $K$, on the system’s sum-rate is illustrated in Fig.~\ref{f10}. It can be observed that the sum-rate improves as $K$ increases. This improvement is due to the enhanced beamforming capability provided by additional transmit antennas at the BS, which enables the system to direct more power toward the intended users. As a result, the achievable sum-rate of the system increases. 

Finally, the impact of varying the upper bound of the channel estimation error parameter, $\delta_m^2$, is illustrated in Fig.~\ref{f11} for both the proposed scheme and its benchmark. As anticipated, the system’s sum-rate declines as $\delta_m^2$ increases for all values of $P_{max}$. This is because larger estimation errors degrade the achievable rates of the common and private data streams for all users, $\tilde{R}_m^o$ and $\tilde{R}_m^p$, defined in \eqref{18} and \eqref{19}, ultimately leading to a reduction in the overall sum-rate. Moreover, since acquiring perfect CSI is nearly impossible due to estimation errors and channel variations, the proposed algorithm demonstrates enhanced robustness against such imperfections. This characteristic makes it particularly well-suited for practical implementation in real-world wireless communication systems, where channel uncertainties are inevitable.

\section{Conclusion}
In this work, we introduced a robust design for a multiuser BD-RIS-empowered RSMA-SWIPT system aimed at enhancing spectral-efficiency, energy-efficiency, coverage, and connectivity in future 6G networks. We jointly optimized the transmit precoding vectors, common rate proportions, power-splitting ratios, and the BD-RIS scattering matrix under imperfect CSI, while accounting for a practical non-linear energy harvesting model. In the proposed system, we developed a robust optimization framework aimed at maximizing the system sum-rate under QoS requirement, power-budget limitations, energy-harvesting constraints, and unitary constraint of scattering matrix, while explicitly considering the worst-case effects of CSI uncertainties. By integrating successive convex approximation, semi-definite programming, and manifold optimization based on the CG method, we computed an efficient solution that explicitly considers all critical system constraints often neglected in prior works. Numerical results validated the effectiveness of our design, showing superior sum-rate performance compared to benchmark schemes.

\section*{\textbf{Appendix A:} Proof of Lemma 1}
 Suppose $\mathbf{P}$ and $\mathbf{Q}$ are two complex matrices, where $\|\mathbf{P}\| \leq 1$. Then, the maximum value of the inner product $\langle \mathbf{P}, \mathbf{Q} \rangle$ corresponds to the dual-norm of $\mathbf{Q}$, i.e.,
\begin{equation}
	\|\mathbf{Q}\|_{dual.} = \operatorname{sup} \{ \operatorname{tr}(\mathbf{Q}^H \mathbf{P}) | \|\mathbf{P}\| \leq 1\} \label{1115}.
\end{equation}
Accordingly, based on \eqref{1115}, the following inequality holds:
\begin{equation}
	\operatorname{tr}(\mathbf{Q}^H \mathbf{P}) \leq \|\mathbf{P}\| \|\mathbf{Q}\|_{dual.}  \label{1116}.
\end{equation}
Then, for Hermitian matrices $\mathbf{Y}$ and $\mathbf{Z}$, we can express the following:
\begin{equation}
	\operatorname{tr}(\mathbf{Y Z}) \leq \|\mathbf{Z}\| \|\mathbf{Y}\|_{dual.} \leq \delta^2 \|\mathbf{Y}\|_{dual.}. \label{117}
\end{equation} 
Thus
\begin{equation}
	\max _{\|\mathbf{Z}\| \leq \delta^2} \operatorname{tr}(\mathbf{Y Z})=\delta^2 \|\mathbf{Y}\|_{dual.}. \label{1118}
\end{equation} 
Additionally, by adopting the spectral norm of $\mathbf{Z}$, we have $\overline{\lambda}_{\max} (\mathbf{Z}) \leq \delta^2$, where $\overline{\lambda}_{\max}$ denotes the maximum eigenvalue of $\mathbf{Z}$. Since the nuclear-norm serves as the dual-norm of the spectral-norm, the dual norm of $\mathbf{Y}$ can be expressed as $\|\mathbf{Y}\|_{\text{dual}} = \|\mathbf{Y}\|_{\text{nuclear}} = \sum_{i} \overline{\lambda}_i$. Furthermore, as $\mathbf{Y}$ is a rank-one PSD matrix, its nuclear norm equals its trace. Therefore, $\|\mathbf{Y}\|_{\text{dual}} = \sum \overline{\lambda} = \operatorname{tr}(\mathbf{Y})$. Consequently, Eq.~\eqref{1118}, can be equivalently expressed as:
\begin{equation}
	\max _{\|\mathbf{Z}\| \leq \delta^2} \operatorname{tr}(\mathbf{Y Z})=\delta^2 \operatorname{tr}(\mathbf{Y}). \label{1119}
\end{equation} 
This completes the proof.

\bibliographystyle{IEEEtran}
\bibliography{Ref}

\end{document}